\documentclass{article}
\usepackage{arxiv}
\usepackage{amsmath}

\usepackage[utf8]{inputenc} % allow utf-8 input
\usepackage[T1]{fontenc}    % use 8-bit T1 fonts
\usepackage{hyperref}       % hyperlinks
\usepackage{url}            % simple URL typesetting
\usepackage{booktabs}       % professional-quality tables
\usepackage{amsfonts}       % blackboard math symbols
\usepackage{nicefrac}       % compact symbols for 1/2, etc.
\usepackage{microtype}      % microtypography
\usepackage{lipsum}		% Can be removed after putting your text content
\usepackage{graphicx}
\usepackage{doi}

\title{Disagreement Concerning Effect-Measure Modification}

\author{ \href{https://orcid.org/0000-0002-5473-7599}{\includegraphics[scale=0.06]{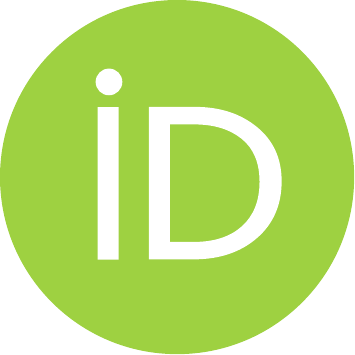}\hspace{1mm}Jake Shannin} \\
	Department of Statistics\\
	University of Florida\\
	Gainesville, FL 32611 \\
	\texttt{jshannin@ufl.edu} \\
	\And
	\href{https://orcid.org/0000-0002-7464-4795}{\includegraphics[scale=0.06]{orcid.pdf}\hspace{1mm}Babette A.~Brumback, Ph.D.} \\
	Department of Biostatistics\\
	University of Florida\\
	Gainesville, FL 32611 \\
	\texttt{brumback@ufl.edu} \\
}

\renewcommand{\shorttitle}{Disagreement Concerning Effect-Measure Modification}

\hypersetup{
pdftitle={The Relative Risks, Cumulative Hazard Ratios, Risk Difference, and Odds Ratio: Disagreement Concerning Effect-Measure Modification},
pdfsubject={Biostatistics: Causal Inference: Effect-Measure Modification},
pdfauthor={Jake Shannin, Babette A.~Brumback},
pdfkeywords={Relative Risks, Cumulative Hazard Ratios, Risk Difference, Odds Ratio, Heterogeneity of Treatment Effect},
}

\begin{document}
\maketitle

\begin{abstract}
	Stratifying factors, like age and gender, can modify the effect of treatments and exposures on risk of a studied outcome. Several effect measures, including the relative risk, hazard ratio, odds ratio, and risk difference, can be used to measure this modification. It is known that choice of effect measure may determine the presence and direction of effect-measure modification. We show that considering the opposite outcome — for example, recovery instead of death — may similarly influence effect-measure modification. In fact, if the relative risk for the studied outcome and the relative risk for the opposite outcome agree about the direction of effect-measure modification, then so will the two cumulative hazard ratios, the risk difference, and the odds ratio. When risks are randomly sampled from the uniform (0,1) distribution, the probability of this happening is 5/6. Disagreement is probable enough that researchers considering one relative risk should also consider the other and further discussion if they disagree. (If possible, researchers should also report estimated risks.) We provide examples through case studies on HCV, COVID-19, and bankruptcy following melanoma treatment.
\end{abstract}

% keywords can be removed
\keywords{Relative Risks, Cumulative Hazard Ratios, Risk Difference, Odds Ratio, Heterogeneity of Treatment Effect}

\section{Introduction}
Many treatments and exposures differently affect different subpopulations. Numerous authors [1-10] have shown that the presence and direction of such effect modification (also called moderation [11]) depends on choice of effect measure — a choice that may appear arbitrary. We will show that the presence and direction of effect-measure modification may also depend on choice between opposite outcomes — another apparently arbitrary choice. For example, effect-measure modification in a study on patients with COVID-19 may depend on whether the outcome is recovery or death. We show that the risk difference (RD $= p2 - p1$) and odds ratio (OR $= \frac{p2(1 - p1)}{p1(1 - p2)}$) are two effect measures immune from this phenomenon, but the relative risk (RR $= \frac{p2}{p1}$) and the cumulative hazard ratio (HR = $\frac{\log(1 - p2)}{\log(1 - p1)}$) are not.

We say two effect measures disagree if they suggest effect modification in opposite directions. Otherwise, they agree. When risks are reported or estimated alongside effect measures, determining agreement or disagreement is straightforward. However, risks are often omitted and occasionally incalculable. Of 222 papers studied by Schwartz et al., 68\% failed to report or estimate risks in the abstract, 35\% failed to report risks anywhere, and 13\% failed to make risks calculable. Papers using adjusted effect measures, such as aOR and aRR, were especially unlikely to report crude risks or estimate adjusted risks. [12] We present methods for determining agreement when risks are unknown.
Existing literature investigates necessary conditions for agreement among RR, RD, and OR. [1] However, except for two articles [13, 14], the literature neglects the potential for disagreement between opposite outcomes for a single effect measure.
Let $p_1$ and $p_2$ denote the proportion of participants in groups 1 and 2 reaching the measured outcome.
We define the other relative risk (RR* = $\frac{1 - p1}{1 - p2}$) to represent the direction of RR modification for the opposite outcome, allowing us to discuss disagreement between opposite outcomes within the existing framework for disagreement between effect measures.
Our literature review also found little discussion of the hazard ratio (hr) surrounding agreement in effect-measure modification.
While the hazard ratio depends on time, we carefully formulate the cumulative hazard ratio to depend only on the risks. The cumulative hazard ratio equals the hazard ratio at all times if the Cox proportional hazards assumption holds. We similarly define the other cumulative hazard ratio, HR* = $\frac{\log(p1)}{\log(p2)}$, to represent HR for the opposite outcome.

Disagreement between the two relative risks, RR and RR*, is not rare: we show in Appendix A that the probability of RR and RR* disagreeing is 1/6 when risks are randomly sampled from the uniform (0,1) distribution. However, we present a theorem that when the two relative risks do agree, all effect measures in our paper (RR, RR*, HR, HR*, RD, OR) — in fact, all effect measure-opposite outcome combinations — agree. This informs our recommendation that researchers present the relative risk for each opposite outcome and further discussion if they disagree. Furthermore, the risks themselves should be included if estimable. Secondary research papers, including meta-analyses, may safely infer effect modification on the HR/HR*/RD/OR scale from the underlying paper’s finding of RR and RR* modification in the same direction.

Our theorem allows substantial dimensionality reduction when testing effect-measure modification across our several effect measures. It suffices to generate simultaneous confidence intervals for just the two relative risk ratios $\frac{p2p3}{p1p4}$ and $\frac{(1 - p1)(1 - p4)}{(1 - p2)(1 - p3)}$ and reject the hypothesis of null effect-measure modification if both intervals lie in the $(<1, <1)$ region or the $(>1, >1)$ region.

The paper is organized as follows: Section \ref{OtherRR} formulates and interprets the other relative risk. Section \ref{Concordant} defines and identifies concordant effect measures, allowing our results to be applied to effect measures beyond the six on which we focus. Section \ref{HazardRatios} formulates and interprets the cumulative hazard ratios, which equal the hazard ratios when the proportional hazards assumption holds. Section \ref{HCV} applies the cumulative hazard ratios to the study of different HCV treatment combinations. Section \ref{Melanoma} explores disagreement in a study on how age modifies the effect of melanoma on risk of bankruptcy. Section \ref{COVID} discusses disagreement regarding how age modifies the effect of health care system on risk of death from COVID-19. Section \ref{MonteCarlo} estimates the probability of effect measures disagreeing when risks are randomly sampled from a uniform distribution. Section \ref{Theorem}, with the help of Appendix B, proves that if the two relative risks agree, then the two cumulative hazard ratios, the risk difference, and the odds ratio agree with them. Appendix A proves that the probability of this happening is 5/6 when risks are randomly sampled from the uniform (0,1) distribution.

\section{The Other Relative Risk}
\label{OtherRR}
One feature of the odds ratio and the risk difference is that if a factor modifies the odds ratio or risk difference for outcome A = 1, then it modifies that effect measure in the same way for outcome A = 0. By modifying two effect measures in ``the same way,” we mean that either both effect measures cross the null (0 for RD, 1 for the other effect measures), or they both move in the same direction, towards or away from the null. We show this in Table \ref{tab:OtherRRtable}.

\begin{table}[ht]
    \centering
    \begin{tabular}{llllll}
    Risk of A = 1 & Control            & Exposure           & RR    & OR    & RD   \\
    Men           & $p1$ = 0.7         & $p2$ = 0.9         & 1.29  & 3.86  & 0.2  \\
    Women         & $p3$ = 0.2         & $p4$ = 0.3         & 1.50  & 1.71  & 0.1  \\
    \hline
    Risk of A = 0 & Control            & Exposure           & RR    & OR    & RD   \\
    Men           & $\tilde{p1}$ = 0.3 & $\tilde{p2}$ = 0.1 & 0.333 & 0.259 & -0.2 \\
    Women         & $\tilde{p3}$ = 0.8 & $\tilde{p4}$ = 0.7 & 0.875 & 0.583 & -0.1 \\
    \end{tabular}
    \caption{Considering participants' risks of the opposite outcome may lead to opposite conclusions if using the relative risk, but not if using the risk difference or odds ratio.}
    \label{tab:OtherRRtable}
\end{table}

In terms of the odds ratio and risk difference, the exposure in Table \ref{tab:OtherRRtable} has a stronger effect on men. Men have a higher OR and RD of A = 1 — and equivalently, a lower OR and RD of A = 0 — than women. This consistency occurs in general: the odds ratio of A = 0 is the reciprocal of the odds ratio of A = 1, and the risk difference of A = 0 is the negative of the risk difference of A = 1.

The relative risk paints a less consistent picture. Women have a RR of A = 1 farther from 1 — but a RR of A = 0 closer to 1 — than men. Take A = 1 as the unfavorable outcome. From the A = 1 RR, we might conclude that the exposure more strongly endangers women. From the A = 0 RR, we might conclude that the exposure threatens men more. From this example, we see that the modifiability of the relative risk depends on choice of outcome, even when choosing between two opposite outcomes — a choice that sometimes appears arbitrary.
Prior research, including Dr. Mindel Sheps' \textit{Shall We Count the Living or the Dead?} [41], discusses this phenomenon but does not, to our knowledge, gauge its frequency or conditions for occurrence.
Later, we will show that this phenomenon — inconsistency in the direction of modification when switching to the opposite outcome — occurs 16.7\% of the time when risks are chosen independently and randomly from the uniform distribution over [0,1]. In fact, this phenomenon does not occur if and only if all six effect measures studied in this paper agree.

One of those effect measures is the other relative risk, RR* = (1 – $p1$)/(1 – $p2$). Let $\tilde{p}1$ = 1 – $p1$ and $\tilde{p}2$ = 1 – $p2$. Then RR* = $\tilde{p}1$/ $\tilde{p}2$, an effect measure defined to reflect what the relative risk would be were we to define risks in terms of the opposite outcome. In the example above, the other relative risk of A = 1 is 3.00 for men and 1.14 for women – values showing the same effect-measure modification by gender as the relative risk of A = 0. Conversely, the other relative risk of A = 0 is 0.778 for men and 0.667 for women, showing the same effect-measure modification by gender as the relative risk of A = 1. In general, $\text{RR*}_{A = 1} = 1/\text{RR}_{A = 0}$, allowing us to study the other relative risk as a proxy to the relative risk for the opposite outcome.

While the other relative risk is rarely invoked explicitly, researchers often choose between reporting risks in terms of two opposite outcomes. Prior research, including similar studies and methodological research, may influence such choices. Government guidelines may also weigh in, such as the 2015 guidelines set by the CDC’s National Center for Health Statistics. These guidelines problematically defined a disparity to grow or shrink if either relative risk increased or decreased a certain amount. [15] Scanlan found contradictions in examples, like vaccination rates and availability of dialysis, where across-the-board improvements shrunk disparities according to one relative risk but grew them according to the other. [13]

Furuya-Kanamori and Suhail suggest choosing the outcome for which $p1$ > $p2$ and $p3$ > $p4$. [14] While they reached an interesting observation that the odds ratio agrees with the more extreme of the relative risks, they did not make a strong argument that the less extreme relative risk always is preferable. Baker and Jackson [44] independently arrive at this conclusion by proposing the generalised relative risk ratio
\begin{equation*}
    \text{GRRR} = \begin{cases}
        \text{RR} - 1 & p2 < p1 \\
        1 - 1/\text{RR*} & p2 \geq p1
    \end{cases}.
\end{equation*}
As they noted, the GRRR is not differentiable and requires a piecewise interpretation. These properties respectively limit the use of the GRRR for methodologists and practitioners.
Instead, we suggest considering both outcomes — or equivalently and more simply, both relative risks RR and RR* — and furthering discussion if the two relative risks disagree.

\subsection{Interpreting the Other Relative Risk}

We can interpret the other relative risk the same way we interpret the relative risk for the opposite outcome. Consider the hypothetical example we presented in Table \ref{tab:OtherRRtable}, for which we compute RR* for each studied gender in Table \ref{tab:OtherRRCalculation}.

\begin{table}[ht]
    \centering
    \begin{tabular}{lllll}
    \textbf{Risk of A = 1} & \textbf{Control} & \textbf{Exposure} & \textbf{RR* Calculation} & \textbf{RR*} \\
    Men                    & $p1$ = 0.7       & $p2$ = 0.9        & $(1-0.7)$/$(1-0.9)$      & 3.00         \\
    Women                  & $p3$ = 0.2       & $p4$ = 0.3        & $(1-0.2)$/$(1-0.3)$      & 1.14    \\    
    \end{tabular}
    \caption{The other relative risk can be calculated directly from the risks of the initially studied outcome. However, it always agrees with the relative risk for the opposite outcome.}
    \label{tab:OtherRRCalculation}
\end{table}

We suggest the following interpretation of RR*, supposing that the initially studied outcome is passing a memory test after drinking tea:
\begin{quote}
    Men drinking decaf tea were 3.00 times as likely to fail the memory test as men drinking caffeinated tea. In contrast, women drinking decaf tea were only 1.14 times as likely to fail the memory test as women drinking caffeinated tea.
\end{quote}
We will see real-world interpretations of RR* in the HCV, Bankruptcy, and COVID-19 case studies (respectively Sections \ref{HCV}, \ref{Melanoma}, and \ref{COVID}).

\section{Concordant Effect Measures}
\label{Concordant}
We define two effect measures as concordant if they always agree. (In Section 2.3, we said that two effect measures disagree when a factor modifies them in different directions. Otherwise, they agree.) This allows us to apply our results involving \{RR, RR*, HR, HR*, RD, OR\} to many other concordant effect measures.

\subsection{The Relative Risk}

RR is concordant with the preventative causal power, CP\textsubscript{p} = ($p1$ – $p2$)/$p1$ = 1 – RR, and Pearl’s probability of necessity, PN = ($p2$ – $p1$)/$p2$ = 1 – 1/RR. We will discuss both in the next subsection. The attributable fraction among the exposed, AF\textsubscript{e} = ($p2$ – $p1$)/$p2$ = 1 – 1/RR, is equivalent to the probability of necessity and therefore also concordant with RR. Similarly, vaccine efficacy, VE = $\frac{p_{control} - p_{vax}}{p_{control}} =$ 1 $-$ RR, [16] is equivalent to the preventative causal power and therefore also concordant with RR. Concordance is transitive; for example, AF\textsubscript{e} and CP\textsubscript{p} are concordant since each is concordant with RR.

\subsection{The Other Relative Risk}

The other relative risk is concordant with generative causal power, CP\textsubscript{g} = ($p2$ – $p1$)/(1 – $p1$) = 1 – 1/RR*. In the words of Glymour, the generative and preventative causal powers are ``brilliant piece[s] of mathematical metaphysics.” When certain assumptions are satisfied, the generative causal power gives the probability that a cause C (in our case, the one present in the $p2$ group and absent in the $p1$ group) would produce an effect E, given that the cause C is absent. [17] The assumptions for this interpretation include:
\begin{itemize}
    \item Let U consist of all causes of E except for C. Then P(C and U) = P(C)P(U). Randomized trials satisfy this assumption (on average).
    \item There is no preventer inhibiting C from causing E. Patricia Cheng initially omitted C from this condition, yielding the following assumption: There is no preventer inhibiting E. Hiddleston prudently qualified this condition to refer specifically to C causing E. [18]
\end{itemize}

With these assumptions, Cheng derived an equivalent to our expression for CP\textsubscript{g}, ($p2$ – $p1$)/(1 – $p1$), from her more general expression for generative causal power. The generative causal power is equal to Pearl’s probability of sufficiency (PS), the probability that a patient not experiencing the cause would have experienced the effect were they to have experienced the cause. [19] (This includes patients who experienced the effect despite not experiencing the cause.) Hence, the probability of sufficiency is also concordant with the other relative risk.

Cheng complemented the generative causal power with the preventative causal power. She defined it for a preventer, e.g. a treatment, preventing the overall set of causes from producing the outcome. Hiddleston generalized this definition, defining preventative causal powers in terms of both a preventer and the cause (or set of causes) it prevents from producing the effect. We prefer to think about preventative causal power as the generative causal power for the opposite outcome. As we discussed in Section 3, the decision of which of two outcomes to measure is a decision often made for convenience. Therefore, if the generative causal power is of general interest, so is the generative causal power of the other outcome: Let $\tilde{p}1$ = 1 – $p1$ and $\tilde{p}2$ = 1 – $p2$. Then CP\textsubscript{g;A = 0} = ($\tilde{p}2$ – $\tilde{p}1$)/(1 – $\tilde{p}1$) = ((1 – $p2$) – (1 – $p1$))/$p1$ = ($p1$ – $p2$)/$p1$ = 1 – RR = CP\textsubscript{p;A=1}, the preventative causal power for the A = 1 outcome.

The preventative causal power is qualitatively aligned with the opposite outcome: when $1 > p2 > p1 > 0$, every effect measure in this paper is greater than its null, except for preventative causal power, which will take a negative value. However, greater preventative causal powers indicate stronger reductions from $p1$ to $p2$, so we say that CP\textsubscript{p} is concordant with the relative risk and probability of necessity, rather than saying that CP\textsubscript{p} always disagrees with RR and PN. In the case of vaccine efficacy, an equivalent to the preventative causal power, we understand a higher vaccine efficacy to reflect a lower relative risk of the disease against which the vaccine protects. The probability of necessity and the preventative causal power have a symmetric relationship: 1 = (1 – PN)(1 – CP\textsubscript{p}).

\subsection{The Risk Difference}

The risk difference is concordant with the number needed to treat (NNT = 1/RD), which gives the number of patients who would need to receive treatment to cure or protect one patient from the disease. The risk difference is equal to, and therefore concordant with, Pearl’s probability of necessity and sufficiency (PNS), which gives the probability that a patient will experience the effect if and only if they experience the cause. [19]

The risk difference is also concordant with — in fact, equal to — the other risk difference, which we define by RD*\textsubscript{A = 1} = –RD\textsubscript{A = 0} = (1 – $p1$) – (1 – $p2$) = $p2$ – $p1$ = RD\textsubscript{A = 1}. Therefore, the risk difference suggests the same conclusion regardless of which of two opposite outcomes is measured. A corollary of RD = RD* is that the risk difference gives equal consideration to the two relative risks:

\begin{equation}
    RD = p_2 - p_1 = (RR - 1)p_1 = \frac{(RR - 1)(RR^* - 1)}{RR \cdot RR^* - 1} = \frac{(RR^* - 1)(RR - 1)}{RR^* \cdot RR - 1} = RD^*
\end{equation}

\subsection{The Odds Ratio}

Edwards found that the odds ratio is concordant with several other measures of association for contingency tables, such as Yule’s Y, Yule’s Q, and the log odds ratio (log OR). [20] The odds ratio is also concordant with, and equal to, the other odds ratio, which we define by OR*\textsubscript{A = 1} = 1/OR\textsubscript{A = 0} = OR\textsubscript{A = 1}. We show this below, letting $\tilde{p}1$ = 1 – $p1$ and $\tilde{p}2$ = 1 – $p2$:

\begin{equation}
    OR^*_{A=1} = \frac{1}{OR_{A = 0}} = \frac{1}{\frac{\Tilde{p_2}(1 - \Tilde{p_1})}{\Tilde{p_1}(1 - \Tilde{p_2})}} = \frac{\Tilde{p_1}(1 - \Tilde{p_2})}{\Tilde{p_2}(1 - \Tilde{p_1})} = \frac{(1 - p_1)p_2}{(1 - p_2)p_1} = OR_{A=1}
\end{equation}

A corollary of OR* = OR is that the odds ratio gives equal treatment to the two relative risks: OR* = (RR*)(RR) = (RR)(RR*) = OR.

\section{The Cumulative Hazard Ratios}
\label{HazardRatios}
\subsection{The Hazard Ratios}

The effect measures we have discussed so far can depend on choice of follow-up period. That is, extending a study’s duration can lead an effect measure to change which stratum it suggests shows a stronger response to exposure or treatment. We define $p_i(t)$ as the risk for stratum $i$, $i = 1, 2$, measured at time $t$. For example, if T is the end of the follow-up period, then $p1$(T) = $p1$. The value of $p_i(0)$ depends on inclusion criteria: If we are studying a treatment for patients with HCV, as in Section \ref{HCV}, we would have $p_i(0) = 1$ since all patients are infected with HCV at the start of the trial. We can now define the hazard rate for each stratum $i$:

\begin{equation}
    h_i(t) = \lim_{\Delta t \rightarrow 0} \frac{p_i(t + \Delta t) - p_i(t)}{(1 - p_i(t)) \Delta t} = \frac{p_i'(t)}{1 - p_i(t)}
\end{equation}

We see that the hazard rate is ill-defined at t = 0 in studies where all patients initially experienced the measured outcome, such as HCV in Section \ref{HCV}. Rather, the hazard rate is meaningful in contexts where patients begin the study not experiencing the outcome and gradually begin experiencing the outcome over the course of the study.

We define the hazard ratio as the ratio of hazard rates for two strata:

\begin{equation}
    \text{hr}(t) = \frac{h_2(t)}{h_1(t)}
\end{equation}

If we took $t = 0$, $p_i(0) = 0$, and $\Delta t = T$ (instead of $\Delta t \rightarrow 0$) when computing the hazard rates, we would get $h_i(t) = p_i/T$, giving a hazard ratio of $p2/p1 = RR$. In this sense, we can think of the hazard ratio as the instantaneous relative risk among patients not already experiencing the outcome. However, we will see that the hazard ratio and relative risk are by no means concordant.

We could apply the hazard rate and ratio to studies where patients all start off experiencing the outcome by considering the opposite outcome — a consideration that sometimes appears arbitrary. Substituting $\Tilde{p}_i$ for $p_i$, we get the hazard rate and ratio for the opposite outcome:

\begin{equation}
    h_i^*(t) = \lim_{\Delta t \rightarrow 0} \frac{\Tilde{p}_i(t + \Delta t) - \Tilde{p}_i(t)}{(1 - \Tilde{p}_i(t)) \Delta t} = -\frac{p_i'(t)}{p_i(t)}
\end{equation}

\begin{equation}
    \text{hr*}(t) = \frac{h_1^*(t)}{h_2^*(t)}
\end{equation}

We will, at times, refer to these as the recovery rate and the recovery ratio. If we took $t = 0$, $p_i(0) = 1$, and $\Delta t = T$ (instead of $\Delta t \rightarrow 0$) when computing the recovery rates, we would get $h_i(t) = (1 - p_i)/T$, giving a recovery ratio of $(1 - p1)/(1 - p2) =$ RR*. Like with RR*, we reciprocated our definition of hr* to ensure qualitative agreement with other effect measures. However, we will see that the recovery ratio and the other relative risk are by no means concordant.

\subsection{The Proportional Hazards Assumption}
Unlike other effect measures in our paper, the hazard ratios (hr and hr*) are functions of time. Not only does this present a challenge to considering agreement between hazard ratios and other effect measures; it also enables self-disagreement, e.g. disagreement between hr($t = 1$) and hr($t = 2$). While these phenomena are interesting, we would like to understand the hazards ratios alongside the other effect measures. One assumption that enables this understanding is the proportional hazards assumption:

\begin{quote}
Let h1($t$) and h2($t$) be the hazard rates for strata 1 and 2. Then we assume there exists a constant $k$ such that h2($t$) = $k$ * h1($t$).
\end{quote}

The most immediate consequence of the proportional hazards assumptions is that it removes the hazard ratio’s dependence of time: $\text{hr(t)} = \frac{h_2(t)}{h_1(t)} = \frac{k h_1(t)}{h_1(t)} = k$, which we assumed to be constant. When the hazards are proportional, we define HR = hr($t$) = $k$ for all times $t$.

For the other hazard ratio, we similarly introduce the proportional recovery assumption:

\begin{quote}
Let h1*($t$) and h2*($t$) be the hazard rates for strata 1 and 2. Then we assume there exists a constant $k$ such that h1*($t$) = $k$ * h2($t$).
\end{quote}

This assumption removes the recovery ratio’s dependence on time: $\text{hr*}(t) = \frac{h_1^*(t)}{h_2^*(t)} = \frac{kh_2^*(t)}{h_2^*(t)} = k$, which we assumed to be constant. When the recovery rates are proportional, we define HR* = hr*($t$) = $k$ for all times $t$.

Conveniently, the proportional hazards assumption is commonly assumed for other statistical procedures, most notably the Cox proportional hazards model. This enables our consideration of the hazard ratios (HR, HR*) alongside our other effect measures (RR, RR*, OR, RD) via an existing assumption. However, there are common situations where these assumptions are unlikely: [21]

\begin{itemize}
    \item Let death be the measured outcome. If the treatment eases symptoms without addressing the underlying disease, patients in the treatment group may experience a favorably low hazard rate early in the study, but not later in the study when patients in the control group have already recovered or died from the disease.
    \item Suppose all patients initially experience the outcome. If some patients in the treatment group improve and then relapse into experiencing the outcome, perhaps as a side effect of the treatment, the recovery rate for their group will likely decrease or even change signs, grossly violating the proportional recovery assumption.
    \item Some treatments accelerate determination of whether the patient will experience the outcome. For example, let the outcome be death and the treatment be surgery. Then patients in the treatment group may experience a high hazard rate during and shortly following surgery but a low hazard rate later into the follow-up period. This may lead to qualitative self-disagreement.
\end{itemize}

\subsection{The Cumulative Hazard Ratios}

To generalize our definition of the time-independent HR and HR* to include cases where the proportional hazards assumption and the proportional recovery assumption do not respectively apply, we define them as the \textit{cumulative} hazard ratios.

More precisely, we define HR $= \frac{\int_0^{t_f}(h_2(t) dt)}{\int_0^{t_f}(h_1(t) dt)}$ as the ratio of cumulative hazard rates and HR* $= \frac{\int_0^{t_f}(h_1^*(t) dt)}{\int_0^{t_f}(h_2^*(t) dt)}$ as the ratio of cumulative recovery rates, where $t_f$ is the duration of the follow-up period for each treatment group. We notably assume equal follow-up periods for the two treatment groups. Appendix C discusses a remedial measure, letting $t_f$ be the duration of the shorter follow-up period, for when the equal follow-up periods assumption fails.

These definitions are compatible with our special case definitions for when hazards or recovery rates are proportional:
\begin{equation}
\text{HR} = \frac{\int_0^{t_f}(h_2(t) dt)}{\int_0^{t_f}(h_1(t) dt)} = \frac{\int_0^{t_f}(kh_1(t) dt)}{\int_0^{t_f}(h_1(t) dt)} = \frac{k\int_0^{t_f}(h_1(t) dt)}{\int_0^{t_f}(h_1(t) dt)} = k
\end{equation}

\begin{equation}
\text{HR*} = \frac{\int_0^{t_f}(h_1^*(t) dt)}{\int_0^{t_f}(h_2^*(t) dt)} = \frac{\int_0^{t_f}(kh_2^*(t) dt)}{\int_0^{t_f}(h_2^*(t) dt)} = \frac{k\int_0^{t_f}(h_2^*(t) dt)}{\int_0^{t_f}(h_2^*(t) dt)} = k
\end{equation}
%use different constant of proportionality in second equation? [I was careful earlier to allow such double use of k.]

Considerable simplification will allow us to compute the cumulative hazard ratios without knowing the time-dependent hazard and recovery rates. First, we rewrite $h_i(t) = \frac{p_i'(t)}{1 - p_i(t)}$ as $\frac{dp_i(t)}{dt} = h_i(t)(1-p_i(t))$. By separation of variables, the solution to this differential equation is $p_i=1-e^{-H_i}$, where $H_i = \int_0^{t_f}(h_i(t) dt)$ is the total hazard for stratum $i$. Note that we used the initial condition $p_i(0) = 0$, reflecting the inclusion criteria that people do not already experience the outcome at the beginning of the follow-up period. Rearranging, we get $H_i = \log{(1 - p_i)}$. This identity allows us to compute the cumulative hazard ratio knowing only the risks: $\text{HR} = \frac{\int_0^{t_f}(h_2(t) dt)}{\int_0^{t_f}(h_1(t) dt)} = \frac{H_2}{H_1} = \frac{\log{(1 - p2)}}{\log{(1 - p1)}}$. %criteria vs criterium

Similarly, we can simplify the other cumulative hazard ratio: Recalling $h_i(t) = -\frac{p_i'(t)}{p_i(t)}$, we define the total recovery $H^*_i = \int_0^{t_f} h_i(t) dt = -\int_0^{t_f} \frac{p_i'(t)}{p_i(t)} dt = \log{p_i(0)} - \log{p_i(t_f)} = -\log{p_i}$ taking $p_i(0) = 1$ by the inclusion criteria that all patients come into the study experiencing the outcome. We can now compute the other cumulative hazard ratio knowing only the risks:

\begin{equation}
\text{HR*} = \frac{\int_0^{t_f}(h_1^*(t) dt)}{\int_0^{t_f}(h_2^*(t) dt)} = \frac{H^*_1}{H^*_2} = \frac{\log{p1}}{\log{p2}}
\end{equation}

Having defined HR and HR* in terms of only $p1$ and $p2$, we can now consider the cumulative hazard ratios alongside the relative risks, odds ratio, and risk difference in our case studies and discussion of agreement and disagreement. For example, we show that if the two relative risks agree, then all six effect measures agree.

\subsection{Interpretation of the Cumulative Hazard Ratios}
%Inclusion criteria  at most one cumulative hazard ratio (when using the log ratio formulae) has an automatic interpretation [optional: move up to here]
In the following HCV case study, we show how only one of the cumulative hazard ratios may have a meaningful interpretation for any given study. A quick way to see which one is to identify the \textit{initial risk} p0. Since none of the patients in the HCV case study started treatment having already reached either endpoint, the proportion of patients who had yet to reach either endpoint on day of treatment $t$ = 0 is p0 = 1. If we had instead defined the risk p as the proportion of patients who \textit{had} reached either endpoint, we would have p0 = 0, suggesting that HR but not HR* would have a natural interpretation.

While in many studies only one cumulative hazard ratio has a meaningful interpretation, it is possible for both to have meaningful interpretations. We will see an example of this in the COVID-19 case study (Section \ref{COVID}). In short, we consider the patients’ risk p of death during their COVID-19 infections, giving p0 = 0 since none of the patients in the study were dead at the beginning of the follow-up period. This accurately suggests a meaningful interpretation for HR. However, it is possible to consider a proportion q of patients who \textit{do not recover} from their COVID-19 infections. Since all patients recover or die from their COVID-19 infections during the follow-up period, q = p. Furthermore, q0 = 1 since none of the patients start having already recovered. As this suggests, we can meaningfully interpret HR* by considering how q($t$) descends from 1 to q = p. Therefore, in the COVID-19 case study, we meaningfully interpret both HR and HR*.

%Comparison of the average hazard ratio and the cumulative hazard ratio (Only a couple sentences)
When the proportional hazards assumption or the proportional recovery rates assumption does not hold, the cumulative hazard ratios will not, in general, equal the average hazard ratios. However, HR and HR* might respectively approximate the average hazard ratios. For example, the hazard rates $h_1(t)=\frac{\sin(t) + 2}{3}$ and $h_2(t)=\frac{\cos(t)+1}{3}$, where $t$ goes from 0 to 10, grossly violate the proportional hazards assumption. Nonetheless, the cumulative hazard ratio HR = 0.433 does not fall too far from the the average hazard ratio $\frac{1}{10}\int_0^{10}(\frac{h_2(t)}{h_1(t)}dt) = 0.486$. In fact, the average hazard ratio, unlike the cumulative hazard ratio, may give undue attention to times during which $h_1(t)$ is near zero. We find the cumulative hazard ratio to be an interesting effect measure even when the proportional hazards assumption does not hold.

\subsection{Case Study: Measuring the Effectiveness of HCV Treatment Combinations}
\label{HCV}
Another common situation in which effect measures may disagree is choice of outcome. Specific effect measures commonly associate with specific outcomes. For example, HIV trials defining participants with less than 50 viral RNA copies per milliliter as having reached the measured outcome (endpoint) typically use risk difference (RD), while trials defining virologic failure as the measured outcome typically use relative risk (RR) or hazard ratio (HR). [3] Different effect measures coupled with different outcomes make summarily interpreting multiple studies especially difficult, as each difference alone is sufficient to lead researchers to different conclusions. In this case study, we review how different outcomes and different effect measures yield diverging conclusions about whether the sofosbuvir/simeprevir combination (sof/sim) or the sofosbuvir/ledipasvir combination (sof/ledi) is more effective at treating patients with the hepatitis C virus (HCV).

%Introduce the endpoints.
Dahari et al. considered two endpoints in their study of how sof/sim and sof/ledi differently affect patients with HCV. Patients first reach endpoint A when their blood has an HCV concentration below 15 IU/ml. They later reach endpoint B when their blood has no detectable HCV. Both endpoints represent definitions of what it means for a patient to have recovered from HCV. This case study will consider two outcomes: outcome A, failure to reach endpoint A within 28 days of treatment, and outcome B, failure to reach endpoint B within 28 days of treatment. These outcomes are coded as one when they occur and zero otherwise. This case study will see how choice of outcome affects several effect measures.

%State data and effect measures.
We define $p1$ = 0.05263 and $p2$ = 0.15000 as the proportions of patients with A = 1 on sof/sim and sof/ledi. [22] This gives RR = 2.850, OR = 3.176, RD = 0.09737, RR* = 1.115, HR = 3.006, and HR* = 1.552. We similarly define $p3$ = 0.26316 and $p4$ = 0.35000 with respect to outcome B. [22] This gives RR = 1.3300, OR = 1.5077, RD = 0.08684, RR* = 1.1336, HR = 1.4106, and HR* = 1.2716.

%Interpret effect measures, especially HR*
Assuming statistical significance, all effect measure-outcome combinations suggest that patients receiving the sof/sim treatment are more likely to recover from their HCV infection within 28 days than patients receiving the sof/ledi treatment. However, we see inconsistencies in the way choice of outcome affects our effect measures. For example, the relative risk sharply falls from 2.85 to 1.33 when switching from outcome A to outcome B, whereas the other relative risk rises slightly from 1.11 to 1.13 under the same change in outcome.
\begin{itemize}
    \item RR: Patients are 2.85 times as likely to have A = 1 on sof/ledi — but only 1.33 times as likely to have B = 1 — as patients on sof/sim.
    \item OR: The odds of A = 1 on sof/ledi are 3.18 times those of patients on sof/sim. In contrast, the odds of B = 1 on sof/ledi are only 1.51 times those of patients on sof/sim.
    \item RD: A patient on sof/ledi is at a 0.0974 greater risk of A = 1 — but only a 0.0868 greater risk of B = 1 — than a patient on sof/sim.
    \item RR*: A patient on sof/sim is 13.4\% more likely to have B = 0 — but only 11.2\% more likely to have A = 0 — than a patient receiving sof/ledi.
    \item HR: In the context of this case study, the cumulative hazard ratios, HR = 3.01 for outcome A and HR = 1.41 for outcome B, have no meaningful interpretations, because we cannot meaningfully define the hazard rates as functions of time. That is, we cannot identify the time during patients’ treatment at which it becomes clear that they will not reach either endpoint.
    \item HR*: However, through regularly testing patients, we can identify the time during patients’ treatment at which they reach either endpoint, giving meaning to the recovery rates. Therefore, the assumption of proportional recovery rates is plausible. If that assumption holds, and since the assumption of equal follow-up periods holds (28 days), patients receiving the sof/sim treatment are 55.2\% more likely to reach endpoint A at any given moment — but only 27.2\% more likely to reach endpoint B at any given moment — than patients receiving the sof/ledi treatment.
\end{itemize}
%Conclude by discussing agreement and disagreement.
Even when two outcomes agree toward identifying which treatment leads to lower risk, different effect measures may disagree as to which outcome suggests a greater difference between the two treatments (or the treatment and a control). In this case study, the other relative risk (RR*) suggested that outcome B, more strongly than outcome A, pointed to the conclusion that sof/sim was more effective than sof/ledi. All other effect measures pointed in the opposite direction, that outcome A more strongly supported the conclusion.

%danger of outcome-effect measure shopping
The numerous effect measure-outcome combinations may enable researchers to cherry-pick the combination with the most extreme value or the lowest P-value. There are several methodological solutions to this problem:
\begin{itemize}
    \item \emph{Pick the combination ahead of the study}. Before beginning a randomized trial or observational endeavor, researchers choose the primary effect measure-outcome combination with which they will analyze results. They might report secondary outcomes and effect measures for purposes of discussion and compatibility with meta-analyses, but not for purposes of determining statistical significance.
    \item \emph{Report the combination standard for the field}. Researchers may base this choice on study design, interpretation of results, or standards in the field. (See the HIV example at the beginning of this case study.)
    \item \emph{Report all combinations}. Researchers may report all relevant effect measure-outcome combinations. While this may prohibit a concise, powerful conclusion, it allows readers to holistically interpret results and maximizes compatibility with meta-analyses.
    \item \emph{Report risks and allow other researchers to compute their preferred effect measure}. Forgoing effect measures entirely, researchers may present the risk of each outcome associated with each treatment, allowing readers to reach their own conclusions. This is nearly as compatible with meta-analyses as reporting all combinations since calculation of effect measures from risks is straightforward. Dahari et al. employed this style of presentation in their HCV study, making no mention of effect measures but presenting all the information this case study needed to compute six and interpret five.
\end{itemize}

\section{Case Study: Young and Old Cancer Patients’ Risks of Bankruptcy}
\label{Melanoma}
\subsection{Introduction}
%The Bankruptcy Abuse Prevention and Consumer Protection Act of 2005 discouraged debtors from filing for bankruptcy. However, it ``had less of an effect on the likelihood of filing among cancer patients than among people without cancer.” \cite{ramsey_washington_2013} %This sentence irrelevant. Remove.
%Is it really worth it to include the graph thing? It more reveals a flaw of the original paper. The discussion is interesting but is it readily communicable? If so, verify/cite the numbers you’re throwing around
%so I know the graph is good for showing HR stuff. But this isn’t in the HR section of the paper (could be moved), so it might turn away readers looking for a basic run of interpretations and agreement/disagreement. Maybe include later in case study?
Melanoma may have a mortality rate of only 0.02\% [23], but it is a financial death sentence much more frequently. Patients with melanoma are on average HR = 2.08 times as likely to file for bankruptcy at any given moment as their matched controls. [24] Ramsey et al. found that $p1$ = 0.00830 of 20-34-year-old patients with melanoma, but only $p2$ = 0.00384 of their matched controls, filed for bankruptcy during an average year of their study. In contrast, only $p3$ = 0.00140 of 80-90-year-old patients with melanoma, and only $p4$ = 0.00045 of their matched controls, filed for bankruptcy during an average year. Assuming that differences are statistically significant and caused by melanoma, we will look at how age modifies effect measures measuring the effect melanoma has on risk of bankruptcy.

\subsection{Effect Measures}
	The relative risk (RR), odds ratio (OR), and cumulative hazard ratio (HR) suggest that melanoma more sharply increases 80-89-year olds' risk of bankruptcy.
\begin{itemize}
	\item RR: 80-89-year-old patients with melanoma are 3.11 times as likely to file for bankruptcy as their matched controls. In contrast, 20-34-year-old patients with melanoma are only 2.16 times as likely to file for bankruptcy as their matched controls.
	\item OR: The odds of 80-89-year-old patients with melanoma filing for bankruptcy are 3.11 times those of their matched controls. In contrast, the odds of 20-34-year-old patients with melanoma filing for bankruptcy are only 2.17 times those of their matched controls.
	\item HR: The assumption of equal follow-up periods holds (1 year). If the proportional hazards assumption also holds, then non-bankrupt 80-89-year-old patients with melanoma are 3.11 times as likely to file for bankruptcy at any given moment as their matched controls. Under the same assumption, non-bankrupt 20-34-year-old patients with melanoma are only 2.17 times as likely to file for bankruptcy at any given moment as their matched controls.
\end{itemize}

The risk difference (RD) and other relative risk (RR*) suggest the opposite conclusion.
\begin{itemize}
	\item RD: The risk of bankruptcy among 20-34-year-old patients with melanoma is 0.00446 higher than the risk of bankruptcy among their matched controls. In contrast, the risk of bankruptcy among 80-89-year-old patients with melanoma is only 0.00095 higher than the risk of bankruptcy among their matched controls.
	\item In terms of the Number Needed to Treat (1/RD), if we relieved 224 20-34-year-old patients with melanoma of its financial effects, we would expect 1 fewer bankruptcy. In contrast, we would have to relieve an estimated 1053 80-89-year-old patients with melanoma from its financial impact to prevent 1 bankruptcy.
	\item RR*: A matched control is 0.45\% (RR* = 1.0045) more likely to avoid bankruptcy than a 20-34-year-old patient with melanoma. In contrast, a matched control is only 0.095\% (RR* = 1.00095) more likely to avoid bankruptcy than an 80-89-year-old patient with melanoma.
\end{itemize}

The other cumulative hazard ratio (HR*) suggests that melanoma equally amplifies time to repayment in 20-34-year-olds and 80-89-year-olds. To interpret HR*, we must make two assumptions that do not exactly fit financial reality:
\begin{itemize}
	\item Proportional recovery rates: We assume that the rate at which indebted patients with melanoma pay off their debt is proportional to that rate for their indebted matched controls.
	\item Equal follow-up periods: We assume that all patients and controls start in debt. During the 1-year follow-up period, we assume that all patients and controls either pay off this debt or declare bankruptcy. This is an exclusive ``or;" we do not account for patients and controls who pay off their debt, accumulate new debt, and subsequently declare bankruptcy within the 1-year period.
\end{itemize}

While not all patients and controls in this study began each studied year in debt, and failure to repay debt does not always equate to bankruptcy, we will continue with an interpretation of HR* to illustrate how HR* may apply to other economic studies of hospital patients, house mortgagees, and car buyers.
\begin{itemize}
	\item Matched controls are 16.1\% (HR* = 1.161) more likely to pay off their debt at any given moment than 20-34-year-old patients with melanoma. Similarly, matched controls are 17.2\% (HR* = 1.172) more likely to pay off their debt at any given moment than 80-89-year-old patients with melanoma.
\end{itemize}

\subsection{Conclusion}
Hospitals often face difficult decisions to stay financially solvent while ensuring that their patients get the care they need. Governments benefit from an understanding of how medical expenses affect citizens’ financial stability since they choose which populations to target with ``treatments” such as Medicare. We include this case study to illustrate how effect-measure modification may be of interest in economics. In causal contexts, effect-measure modification is the study of how a modifier affects the extent to which an exposure causes a disease. In this case study, the modifier is age, the exposure is melanoma, and the ``disease” is bankruptcy. Financial events — bankruptcy, repayment, making a purchase, clicking on an advertisement, selecting a contractor, entering a recession — are meaningfully compatible with effect measures. We hope that future research incorporates effect-measure modification within econometrics.

\section{Case Study: COVID-19 Mortality and Country of Treatment}
\label{COVID}
\subsection{Introduction}

The risk of death in patients with COVID-19 depends heavily on many factors including (a) their age [25] and (b) the relative prevalence of COVID-19 in their healthcare system, relative to that system's capacity. In this case study, we use different effect measures to investigate how the age of patients with COVID-19 modifies the effect their healthcare system has on their risk of death.

For purposes of this case study, we will neglect confounders, i.e., mutual causes of COVID-19 mortality and relative prevalence. Such variables include the rate of testing: increased testing decreases the measured death rate of COVID-19 by revealing asymptomatic and weakly symptomatic cases. [26] Increased testing also decreases the relative prevalence of COVID-19, since countries with increased testing generally detect COVID-19 outbreaks in time to implement appropriate policy actions to prevent the outbreak from overwhelming their health care systems. [27] Thus, increased testing partially "explains away" the strong association between COVID-19 mortality and relative prevalence. We therefore intend this case study as an illustrative example of how effect-measure modification changes from one effect measure to another.

\subsection{Effect Measures}

Humans in their forties generally experience a below average risk of dying from COVID-19. Italian 40-49-year-olds are no exception with a death rate of $p1$ = 0.9\% as of June 3, 2020. [28] Mexican 40-49-year-olds are not as fortunate with a death rate of $p2$ = 7.5\% as of June 3, 2020. [29] Among 60-69-year-olds, Italians and Mexicans have respective Case Fatality Rates of $p3$ = 10.6\% and $p4$ = 25.3\% (RR = 2.39). While COVID-19 overwhelmed both countries' healthcare systems, it caught Mexico particularly unprepared [30], at least partially explaining these disparate death rates. Other explanatory variables include Mexico’s increased absolute prevalence and accelerated onset of preexisting conditions that increase the risk of death from COVID-19. [30] We will look at how age modifies each of our effect measures.

The relative risk (RR), odds ratio (OR), and cumulative hazard ratio (HR) find the disparity between Mexican and Italian 40-49-year-olds more alarming than that disparity among 60-69-year-olds.
\begin{itemize}
	\item RR: A 40-49-year-old person from Mexico with COVID-19 is 8.33 times as likely to die as a 40-49-year-old person from Italy with COVID-19. In contrast, a 60-69-year-old person from Mexico with COVID-19 is only 2.39 times as likely to die as their Italian counterpart.
	\item OR: The odds of Mexican 40-49-year-olds with COVID-19 dying are 8.93 times those of Italian 40-49-year-olds. In contrast, the odds of Mexican 60-69-year-olds with COVID-19 of dying are only 2.86 times those of their Italian counterparts.
	\item HR: Assuming proportional hazards and equal follow-up periods, Mexican 40-49-year-olds with COVID-19 are 8.62 times as likely to die at any given moment as Italian 40-49-year-olds with COVID-19. Under the same assumptions, Mexican 60-69-year-olds with COVID-19 are only 2.60 times as likely to die at any given moment as their Italian counterparts.
\end{itemize}
These effect measures may lead stakeholders to conclude that countries with underprepared healthcare systems should focus on middle-age patients whose deaths are possible but typically preventable, rather than on older patients who have a substantial chance of dying even if prioritized for treatment.

The risk difference (RD) and other relative risk (RR*) yield the opposite conclusion.
\begin{itemize}
	\item RD: The risk of death among 60-69-year-old Mexicans with COVID-19 is 0.147 higher than the risk of death among 60-69-year-old. In contrast, the risk of death among 40-49-year-old Mexicans with COVID-19 is only 0.066 higher than that risk among their Italian counterparts.
	\item In terms of the Number Needed to Treat (1/RD) and using our assumption of causation, if 6.8 (1/0.147) 60-69-year-old Mexicans with COVID-19 were instead treated under Italy's healthcare system, we would expect 1 fewer death. In contrast, we would have to treat an estimated 15.2 40-49-year-old Mexicans with COVID-19 under Italy's healthcare system to save 1 life.
	\item RR*: A 60-69-year-old person from Italy with COVID-19 is 19.7\% (RR* = 1.197) more likely to survive infection than a 60-69-year-old person from Mexico with COVID-19. In contrast, a 40-49-year-old person from Italy with COVID-19 is 7.1\% (RR* = 1.071) more likely to survive infection than a 40-49-year-old person from Mexico with COVID-19.
\end{itemize}
Risk difference is arguably the effect measure most suitable for identifying which subpopulation would benefit the most from treatment. [13, 31-33]

The other cumulative hazard ratio (HR*) does not suggest that age substantially modifies the effect of healthcare system on mortality from COVID-19: Assuming proportional recovery rates and equal follow-up periods during which all patients recover or die from COVID-19, 40-49-year-old people from Italy are 81\% more likely to recover from COVID-19 at any given moment than 40-49-year-old people from Mexico with COVID-19. Similarly, 60-69-year-old people from Italy are 63\% more likely to recover from COVID-19 at any given moment than 60-69-year-old people from Mexico under the same assumptions.

\subsection{Conclusion}

The COVID-19 pandemic caught healthcare systems unprepared, requiring them to choose which subpopulations to treat with limited resources. Data detailing these subpopulations’ risks of death from COVID-19 with and without treatment inform such decisions. Our case study suggests that the effect measure used to compare these data may determine this decision: Mexico may target treatment toward 40-49-year-old patients with COVID-19 if they compare data with the relative risk, odds ratio, or cumulative hazard ratio; alternatively, Mexico may prioritize treating 60-69-year-old patients with COVID-19 after comparing data with the risk difference or the other relative risk (i.e., the relative risk of surviving COVID-19). This case study serves as an example of meaningful disagreement between effect measures. The effect measures we studied differed between strata substantially (RR: 8.33 vs. 2.39; RD: 0.066 vs. 0.147), showing that disagreement does not only occur in cases where, perhaps, RR is slightly higher for 40-49-year-olds and RD is slightly higher for 60-69-year-olds. A large difference in one effect measure does not guarantee that other effect measures agree.

\section{Monte Carlo Simulation}
\label{MonteCarlo}
Through case studies, we have shown that effect measures can disagree in real-world examples. But just how common is this phenomenon? To get a general sense for the prevalence of agreement and disagreement, we compute the probability that pairs and sets of effect measures agree when risks are randomly sampled from the unit interval. We also compute the probability of agreement when risks are randomly selected in $0 < p_i < 0.1$ to gauge the prevalence of disagreement in the study of rare diseases. [1]

\subsection{Methods}

To compute these probabilities, we performed Monte Carlo integration in Java. In each of one million trials, we randomly sampled risks $p1$, $p2$, $p3$, and $p4$ from the uniform (0,1) distribution. We computed our six effect measures, \{RR, RR*, HR, HR*, RD, OR\}, for each the $\{p1, p2\}$ stratum and the $\{p3, p4\}$ stratum. We say a set of effect measures disagrees if any pair disagrees. We then determined and recorded, for each of the 64 subsets of our six effect measures, whether that subset agreed or disagreed. We estimate the probability of a given set of effect measures agreeing by the proportion of the one million trials in which that set agreed. We repeated this process for rare diseases by sampling risks from the uniform (0, 0.1) distribution.

\subsection{Results}

We present our estimates of the probabilities of sets of effect measures agreeing in the below 6-way Venn Diagrams [34]:

\begin{figure}[htbp]
\centering
\includegraphics[scale = 0.425]{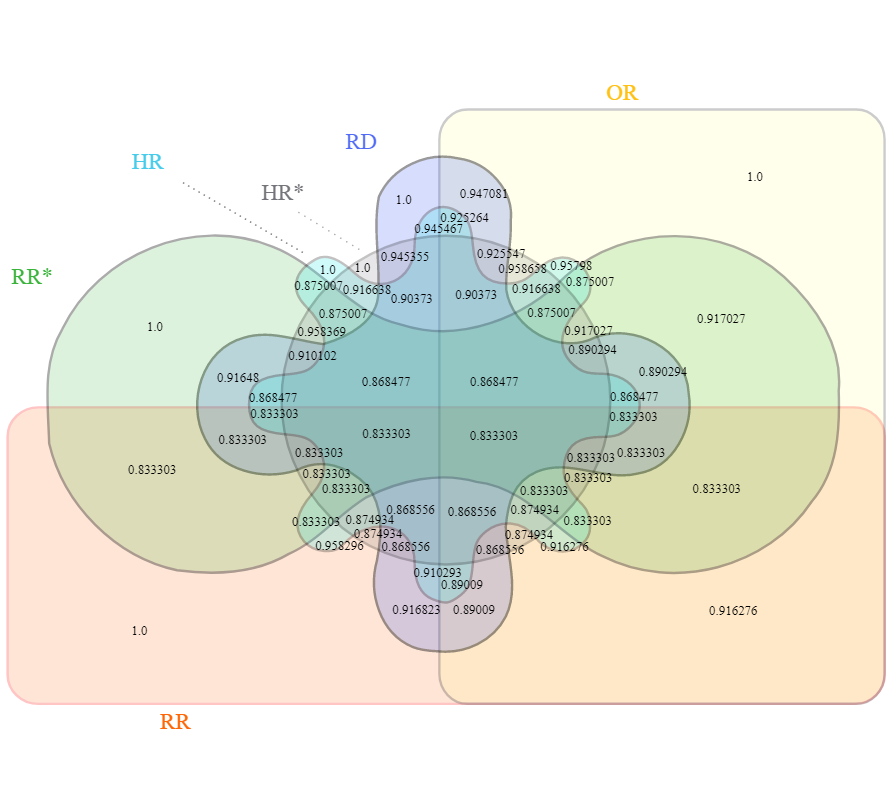}
\caption{We present the approximate probability of agreement for various sets of effect measures when risks are randomly sampled from the (0,1) distribution, imitating risks observed in the study of common outcomes. Probabilities do not add to 1 because they do not represent mutually exclusive events. For example, if all six effect measures agree (probability 0.833303), then every subset of those effect measures also agrees, so all probabilities are at least 0.833303.}
\label{CommonVenn}
\end{figure}

\begin{figure}[htbp]
\centering
\includegraphics[scale = 0.425]{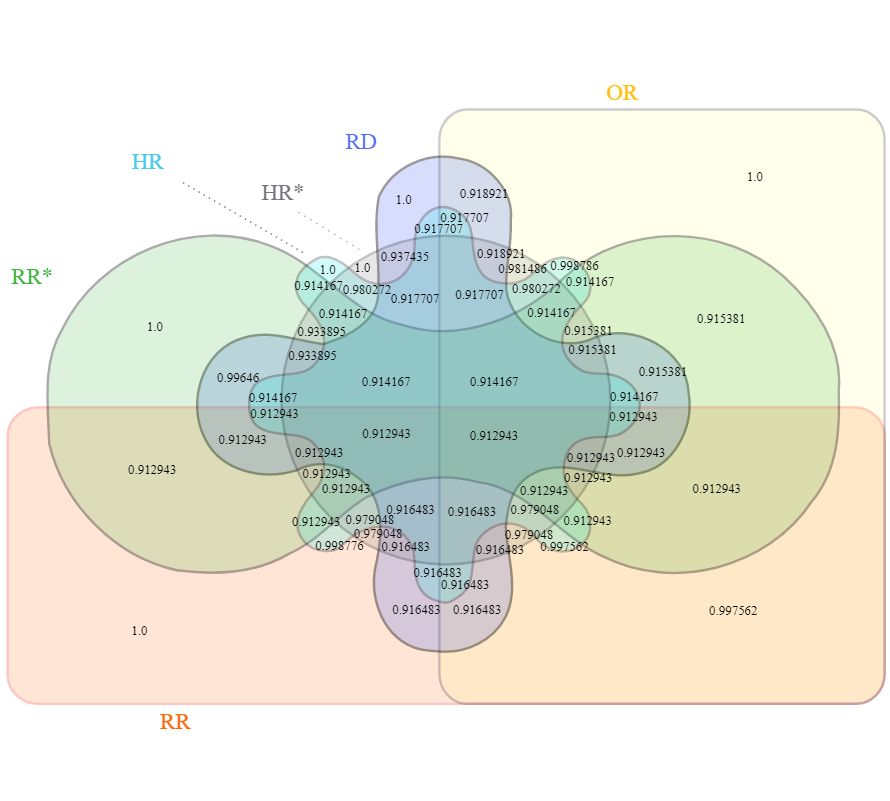}
\caption{We present the approximate probability of agreement for various sets of effect measures when risks are randomly sampled from the (0,0.1) distribution, imitating risks observed in the study of rare outcomes.}
\label{RareVenn}
\end{figure}

\subsection{Analysis}

The most striking pattern in Figure \ref{CommonVenn} is the repeated occurrence of 0.833303. This is our estimate of the probability that all six effect measures agree when risks are randomly sampled from the unit interval. (In Appendix A, we show that the exact probability is $\frac{5}{6}$.) But this is also our estimate that numerous subsets of those effect measures agree. The minimal such subset is the pair of relative risks: 0.833303 is our estimate of the probability that the two relative risks agree. From this equality of probabilities, we conjecture that all six effect measures agree if and only if the two relative risks agree. We prove this in the following section.

Suppose there were many studies in which risks were independent and uniformly distributed over the unit interval and in which all relevant inequalities reach significance. Then this probability, 0.833303, suggests that about one in six ($1 - 0.833303$) of those studies, if they used the relative risk to assess effect-measure modification, would reach the opposite conclusion were they to have instead considered the opposite outcome — a consideration that sometimes appears arbitrary. Since neither assumption is reasonable [45], we do not claim that one-sixth of all RR modification studies would reach the opposite conclusion were the outcome oppositely codified. (In Appendix D, we redo our simulation by sampling $p2$ and $p4$ from tent distributions that respectively depend on $p1$ and $p3$.)

However, we do suspect that this proportion is high enough to raise concerns about focusing on only the relative risk and only one outcome. As we outlined in Section \ref{HCV}, one alternative is to consider all effect measure-outcome combinations, or more succinctly, all effect measures including those representing the opposite outcome (RR*, HR*). Another option is to present the risks themselves, allowing other researchers to compute the effect measures in which they are interested.

From our result that all six effect measures agree exactly when the two relative risks agree, we suggest another alternative: Researchers can initially consider the two relative risks. If they agree, the conclusion they suggest can be taken to apply to other effect measures. If they disagree, the researchers can discuss them alongside other effect measures and consider which effect measures are most applicable or standard for their topic. For example, risk difference is commonly used in choosing which population to prioritize for scarce treatment. [13, 31-33]

There are several more conjectures we draw from repeated Monte Carlo probabilities:
\begin{itemize}
    \item If HR and HR* agree, so does OR.
    \item If HR and RR* agree, so does HR*.
    \item If OR and RR* agree, so does HR*.
    \item If HR* and RR agree, so does HR.
    \item If RR and OR agree, so does HR.
\end{itemize}
While we will not prove these conjectures, our large number of trials (1 million) allows us to be confident that they are safe to use in applied contexts.

We now turn our attention to Figure \ref{RareVenn}. We see that agreement is a more common phenomenon in the context of rare diseases: all effect measures agree with estimated probability 0.912943. Unlike in Figure \ref{CommonVenn}, this proportion is just slightly below the probability of the odds ratio and risk difference agreeing: 0.915381. This follows from how for rare diseases, the odds ratio approximates the relative risk (agreement probability 0.997562), and the risk difference approximates the other relative risk (agreement probability 0.996460). For rare diseases, another consequence of RD approximating RR* is that if RR and RD agree, then almost surely will all six effect measures agree.

We find this schism between the odds ratio and the risk difference concerning in the context of case-control studies, which are commonplace in the study of rare diseases and, by design, report the odds ratio. Since the risk difference is more appropriate for choosing which population to target treatment toward, we suggest that case-control studies also present RD when a known sampling fraction allows estimation of the risks $p1$ and $p2$. [35]

\section{Theorem: If Both Relative Risks Agree, then so do Both Hazard Ratios, the Risk Difference, and the Odds Ratio.}
\label{Theorem}
In Section \ref{MonteCarlo}, the two relative risks (RR and RR*) agreed in the exact same number of trials, 833303 out of 1000000, as the entire set of effect measures \{RR, RR*, HR, HR*, RD, OR\}. From this we conjecture that the relative risks agree if and only if all six effect measures agree. We can prove this in two ways: algebraically and analytically. In this section, we provide the algebraic proof. The analytic proof, partially presented in Appendix A, is more technical than the algebraic proof but offers a framework useful for future proofs, such as our proof in Appendix A that all six effect measures agree with probability $\frac{5}{6}$ when risks are randomly sampled from the uniform (0,1) distribution.

To ensure all effect measures are defined, we restrict the risks $p1, p2, p3,$ and $p4$ to the open unit interval. If at least two risks are 0 or at least two risks are 1, the presence and direction of effect modification are clear without the use of effect measures. If exactly one risk is 0 and at most one risk is 1, or vice versa, then it is feasible to apply our theorem by defining otherwise-undefined effect measures using appropriate one-sided limits. For example, if $p1 = 0$ and $p2 = 0.3$, then we may define RR = $\lim_{p1 \rightarrow 0^+} \frac{0.3}{p1}$ = OR = HR = HR* = $\infty$.

\subsection{Qualitative Effect Modification: If either \texorpdfstring{$p1 < p2$}{p1 < p2} or \texorpdfstring{$p3 < p4$}{p3 < p4} but not both, then all six effect measures agree.}
\label{Qualitative}
Relabelling strata as necessary, suppose that $p1 < p2$ and $p3 \geq p4$. Let RR\textsubscript{P} denote the relative risk for the $(p1, p2)$ stratum, and let RR\textsubscript{Q} denote the relative risk for the $(p3, p4)$ stratum. We similarly define this subscript notation for other effect measures. Then
\begin{itemize}
    \item $\text{RR}_P > 1$ and $\text{RR}_Q \leq 1$,
    \item $\text{RR*}_P > 1$ and $\text{RR*}_Q \leq 1$,
    \item $\text{HR}_P > 1$ and $\text{HR}_Q \leq 1$,
    \item $\text{HR*}_P > 1$ and $\text{HR*}_Q \leq 1$,
    \item $\text{RD}_P > 0$ and $\text{RD}_Q \leq 0$, and
    \item $\text{OR}_P > 1$ and $\text{OR}_Q \leq 1$.
\end{itemize}
Hence all six effect measures agree as desired.

\subsection{If RR and RR* agree, then HR* agrees with them.}
\label{RR/RR*/HR*}
If $p1$ $<$ $p2$ or $p3$ $<$ $p4$ but not both, we have qualitative effect modification, so by Section \ref{Qualitative}, \{RR, RR*, HR*\} agrees as desired. Otherwise, we can relabel treatment groups and strata as necessary so that $1 < \frac{p2}{p1} < \frac{p4}{p3}$. Suppose that RR and RR* agree, giving that $1 < \frac{1 - p1}{1 - p2} < \frac{1 - p3}{1 - p4}$. We will show that $\frac{\log p1}{\log p2} < \frac{\log p3}{\log p4}$ in the two below cases. Note that we write each case as a standalone proposition; i.e., we rewrite aforementioned suppositions that we use in each case's proof.\\

\underline{Suppose $p4$ $>$ $p2$ $>$ $p1$ and $p4$ $>$ $p3$ $>$ $p1$. Then RR and HR* agree.}\\
%I don't think 1 < RR is used even though we're allowed it
Taking the logarithm of both side of $\frac{p2}{p1} < \frac{p4}{p3}$ gives that $\log p2 - \log p1 < \log p4 - \log p3$, from which it follows that $\frac{\log p1}{\log p2} < \frac{\frac{\log p3}{\log p4} - 1}{\frac{\log p2}{\log p4}} + 1$. And since $p4 > p2$, we have $\frac{\frac{\log p3}{\log p4} - 1}{\frac{\log p2}{\log p4}} + 1 < \frac{\log p3}{\log p4}$. Transitively, $\frac{\log p1}{\log p2} < \frac{\log p3}{\log p4}$, so RR and HR* agree as desired.\\

\underline{Suppose $p4 < p2 < 1 < \frac{1 - p1}{1 - p2} < \frac{1 - p3}{1 - p4}$. Then RR* and HR* agree.}\\
For $i \in \{1, 2, 3, 4\}$, let $\tilde{p_i} = 1 - p_i$. Then
\begin{equation}
    \frac{\log p1}{\log p2} = \frac{\log(1 - \tilde{p}_2\text{RR*}_P)}{\log(1 - \tilde{p}_2)} < \frac{\log(1 - \tilde{p}_4\text{RR*}_P)}{\log(1 - \tilde{p}_4)} < \frac{\log(1 - \tilde{p}_4\text{RR*}_Q)}{\log(1 - \tilde{p}_4)} = \frac{\log p3}{\log p4}
\end{equation}
with inequalities justified in Appendix B. Since RR*\textsubscript{P} $<$ RR*\textsubscript{Q} and HR*\textsubscript{P} $<$ HR*\textsubscript{Q}, we have that RR* and HR* agree.\\

In the remaining case that $p2$ = $p4$, all six effect measures agree, including RR, RR*, and HR*. Hence if the two relative risks agree, then the other cumulative hazard ratio agrees with them as desired.

\subsection{If RR and RR* agree, then HR agrees with them.}

Conceptually: The relative risks are concordant with each other for the opposite outcome, so if RR and RR* agree for one outcome, they also agree for the opposite outcome. By the previous subsection, this implies that HR* for the opposite outcome agrees. But HR* for the opposite outcome is concordant with HR, so if RR and RR* agree, then HR agrees with them as desired. \\

Algebraically: We showed in Section \ref{RR/RR*/HR*} that for any $w,x,y,z \in (0,1)$, if $\frac{x}{w} > \frac{z}{y}$ and $\frac{1 - w}{1 - x} > \frac{1 - y}{1 - z}$, then $\frac{\log w}{\log x} > \frac{\log y}{\log z}$. Let $w = 1 - p2, x = 1 - p1, y = 1 - p4$, and $z = 1 - p3$. Then the above gives that if $\frac{1 - p1}{1 - p2} > \frac{1 - p3}{1 - p4}$ and $\frac{p2}{p1} > \frac{p4}{p3}$, then $\frac{\log(1 - p2)}{\log(1 - p1)} > \frac{\log(1 - p4)}{\log(1 - p3)}$. Similarly, if $\frac{1 - p1}{1 - p2} < \frac{1 - p3}{1 - p4}$ and $\frac{p2}{p1} < \frac{p4}{p3}$, then $\frac{\log(1 - p2)}{\log(1 - p1)} < \frac{\log(1 - p4)}{\log(1 - p3)}$. Therefore if the two relative risks agree, then the cumulative hazard ratio agrees with them as desired.

\subsection{If RR and RR* agree, then RD agrees with them.}

We begin by presenting two standalone sufficient conditions for agreement:

\underline{If RR*\textsubscript{P} $<$ RR*\textsubscript{Q} and $p3$ $\leq$ $p1$, then RR* and RD agree.}\\
Suppose that $\frac{1 - p1}{1 - p2} < \frac{1 - p3}{1 - p4}$. Equivalently, $\frac{p3 - p4}{1 - p3} < \frac{p1 - p2}{1 - p1}$. Therefore $p4 - p3 > \frac{(p2 - p1)(1 - p3)}{1 - p1} \geq p2 - p1$ by our assumption that $p3$ $\leq$ $p1$. Since RD\textsubscript{P} $<$ RD\textsubscript{Q}, the other relative risk and the risk difference agree.\\

\underline{If RR\textsubscript{P} $<$ RR\textsubscript{Q} and $p3$ $\geq$ $p1$, then RR and RD agree.}\\
Suppose that $\frac{p2}{p1} < \frac{p4}{p3}$. Then $\frac{p2 - p1}{p1} < \frac{p4 - p3}{p3}$. Since $p3 \geq p1$, we have $p4 - p3 > (p2 - p1)\frac{p3}{p1} \geq p2 - p1$. Hence RD\textsubscript{P} $<$ RD\textsubscript{Q}, so the relative risk and the risk difference agree.\\

Suppose that RR and RR* agree. If there is qualitative effect modification, all effect measures, including RD, agree. Otherwise, we relabel strata as necessary so that RR\textsubscript{P} $<$ RR\textsubscript{Q} and RR*\textsubscript{P} $<$ RR*\textsubscript{Q}. Then one of the two above results applies, i.e., the risk difference must always agree with one of the relative risks. We conclude that if the two relative risks agree, then the risk difference must agree with them as desired.

\subsection{If RR and RR* agree, then OR agrees with them.}

This follows from the odds ratio being the product of the two relative risks: Suppose that $\frac{p2}{p1} < \frac{p4}{p3}$ and $\frac{1 - p1}{1 - p2} < \frac{1 - p3}{1 - p4}$. Multiplying these inequalities, we get that $\frac{p2(1 - p1)}{p1(1 - p2)} < \frac{p4(1 - p3)}{p3(1 - p4)}$. Since RR\textsubscript{P} $<$ RR\textsubscript{Q} and RR*\textsubscript{P} $<$ RR*\textsubscript{Q} together imply OR\textsubscript{P} $<$ OR\textsubscript{Q}, we conclude that if the two relative risks agree, then the odds ratio agrees with them as desired.

\subsection{Conclusion}

From the preceding sections, we conclude that if the two relative risks agree, then so must the set of all six of our effect measures. Along the way, we found several sufficient conditions for agreement between effect measures:
\begin{itemize}
    \item If $p4 > p2 > p1$ and $p4 > p3 > p1$ are both true or both false, then RR and HR* agree. Otherwise, RR* and HR* agree.
    \item If $p4 < p2$ and $\frac{1 - p1}{1 - p2} < \frac{1 - p3}{1 - p4}$ are both true or both false, then RR* and HR* agree. Otherwise, RR and HR* agree.
    \item If RR*\textsubscript{P} $<$ RR*\textsubscript{Q} and $p3$ $\leq$ $p1$ are both true or both false, then RR* and RD agree. Otherwise, RR and RD agree.
    \item If RR\textsubscript{P} $<$ RR\textsubscript{Q} and $p3$ $\geq$ $p1$ are both true or both false, then RR and RD agree. Otherwise, RR* and RD agree.
\end{itemize}

\section{Discussion}
\label{Discussion}
Researchers use a variety of effect measures to quantify health disparities and other instances of effect-measure modification. As we saw in our case studies on COVID-19, HCV, and outpatient bankruptcy, choice of effect measure can determine the existence and direction of modification when risks increase in both strata or decrease in both strata. As in previous literature [1], we defined two effect measures to \textit{disagree} if the stratifying factor (e.g., gender) modifies the two effect measures in opposite directions. Otherwise, they \textit{agree}. Two effect measures that always agree are \textit{concordant}.

Moreover, researchers often choose between opposite outcomes, such as recovery and death in a study where all patients recover or die. We showed that the risk difference (RD) and the odds ratio (OR) are unaffected (for purposes of agreement) by this choice, but the relative risk (RR) and the cumulative hazard ratio (HR) may suggest opposite conclusions. In fact, if the relative risk for one outcome and the relative risk for the other outcome suggest effect-measure modification in the same direction, then so will all other aforementioned effect measures for either outcome.

\subsection{Proofs}
We defined RR* as the reciprocal of the relative risk for the opposite outcome and HR* as the reciprocal of the cumulative hazard ratio for the opposite outcome. For any risks $p1$, $p2$, $p3$, and $p4$ between 0 and 1, we showed that HR, HR*, RD, and OR each agree with at least one of the relative risks. As a result, if the two relative risks RR and RR* agree, then so does the entire set of our effect measures \{RR, RR*, HR, HR*, RD, OR\}. We proved the entire theorem algebraically, uncovering several sufficient conditions for agreement along the way. We also used the intermediate value theorem to show that the risk difference agrees with at least one of the relative risks. The latter proof provides the framework we used to prove that all our effect measures agree with probability 5/6 when risks are randomly sampled from the uniform (0,1) distribution.

\subsection{Simulation}
While our Monte Carlo simulation served primarily to foretell our theorem, it uncovered several more agreement relations. For example, if risks are randomly sampled from the uniform (0, 0.1) distribution — modeling the use of effect measures to study rare diseases — and the relative risk and risk difference agree, then our theorem approximately applies: all six of our effect measures agreed in 912943 of the 916483 trials in which RR and RD agreed. In only 91.9\% of the rare disease trials did the odds ratio and risk difference agree, a concerningly low probability given that case-control studies often yield only the odds ratio, while the risk difference is the effect measure most suitable for determining which population to prioritize for scarce treatment. [13, 31-33]

\subsection{Case Studies}
Throughout our paper, we considered the choice between two opposite outcomes — a choice often made for convenience. In our HCV case study, we additionally considered the choice between related outcomes: having a low bloodborne HCV concentration and having no detectable bloodborne HCV whatsoever. Assuming statistical significance, we showed that choice between these two outcomes modified each of our effect measures, but not all in the same direction: RR dropped from 2.85 to 1.33 when switching from the former outcome to the latter, while the RR* gently increased from 1.115 to 1.134.

In our case study on how age modifies the effect of melanoma on risk of bankruptcy, we put one foot in finance while keeping the other in medicine. This case study dispelled the notion that when effect measures disagree, they only do so “slightly:” the risk difference suggested that the younger age group sees a much sharper effect of melanoma on risk of bankruptcy (RD = 0.00446 compared to RD = 0.00095 for the older age group), while the relative risk suggested that the older age group sees a sharper increase in risk (RR = 3.11 compared to RR = 2.16 for the younger age group).

Finally, our case study on COVID-19 provided an example of effect measures disagreeing in a critical public health context. We showed that age modifies the effect of health care system on risk of death from COVID-19 in opposite directions for different effect measures. As a result, choice of effect measure could impact which age groups receive prioritized treatment. We recommend the risk difference for this context due to its concordance with the number needed to treat. [13, 31-33]

\subsection{Existence of effect-measure modification}

Our paper focuses mainly on the direction of effect-measure modification, as much is already known about existence. For example, if relative risk modification does not occur, and all risks are distinct and strictly between 0 and 1, then effect-measure modification necessarily occurs on the RR*, HR, HR*, RD, and OR scales. However, it is not quite true that if effect-measure modification does not occur on any one of these scales, then it must occur on the other five. For example, if $p1 + p4 = p2 + p3 = 1$, then effect-measure modification occurs on neither the RD nor the OR scale.

\subsection{Shortcomings of existing literature}

Some authors have written about the potential for the presence and direction of relative risk modification to depend on which of two opposite outcomes are chosen. Following an investigation of 551 meta-analyses, Deeks concluded that the two relative risks "are best considered as separate models." [42] More recently, Scanlan provided an example where the U.S. government’s definition of disparity would classify a racial disparity in vaccination rate as having both increased and decreased in the same period, depending on whether risks were defined by the vaccination rate or the proportion unvaccinated. [13] Our paper shows that the probability of this phenomenon may be concerningly high: 1/6 were risks randomly sampled from the uniform (0,1) distribution. While these prior papers show the prudency in reporting both relative risks, our paper goes further by showing that agreement between the two relative risks is sufficient for agreement between all six effect measures we study.

\subsection{How to apply}
Our findings are of interest to researchers choosing between effect measures and opposite outcomes and to researchers performing meta-analyses over literature employing varying effect measures and outcome codifications. In some fields, there is a standard effect measure-outcome combination. In some studies, the purpose of the study informs the effect measure-outcome choice: a study recommending a population for prioritized COVID-19 vaccination may employ the risk difference to save the most lives. In contexts where there is no clear choice, we recommend that researchers report both relative risks. If they agree, our theorem shows that the studied factor also modifies HR, HR*, RD, and OR in the same direction. For example, a meta-analysis of studies testing for risk difference modification could include a study that showed relative risk modification for each of two opposite outcomes.

\subsubsection{Bivariate delta method}
Brumback and Berg suggested the multivariate delta method to test the alternative hypothesis that a factor modifies the relative risk, risk difference, and odds ratio in the same direction. [1] This method involves considering a joint distribution with a dimension for each of the three effect measures. We improve on this recommendation, increasing the strength of the alternative hypothesis and reducing the dimensionality of the applicable joint distribution: we suggest using the bivariate delta method to test the alternative hypothesis that a factor modifies both relative risks, both cumulative hazard ratios, the risk difference, and the odds ratio in the same direction. By our theorem, it suffices to consider the joint distribution of just the two relative risk ratios $\frac{p2p3}{p1p4} \text{ and } \frac{(1 - p1)(1 - p4)}{(1 - p2)(1 - p3)}$. We reject the null hypothesis if the 100(1 – $\alpha$)\% simultaneous confidence region for the relative risk ratios lies completely within the (>1, >1) region or the (<1, <1) region.

\subsection{Future Research}
There are several other topics in effect-measure modification to which future research might apply our research.
\begin{itemize}
    \item Gilbert et al. adapt survivor average causal effect (SACE) analysis to principal surrogate (PS) analysis on the HVTN 505 HIV-1 vaccine trial. Their analysis found qualitative vaccine efficacy modification by a post-randomization biomarker. [36] Future research may adapt SACE to PS in the context of effect measures besides vaccine efficacy (which is concordant with the relative risk). Furthering our consideration of the opposite outcome, future research could formulate the ``other'' survivor average causal effect (SACE*) to be the average causal effect in participants who would be non-survivors (e.g., who would experience HIV infection) regardless of assignment to the control or treatment group.
	\item Dahabreh et al. discuss the limitations of considering ``one variable at a time” when choosing a treatment for a patient with multiple risk factors. [10] Future research could explore scenarios where multiple factors modify some effect measures but not others.
	\item Conversely, Rauch et al. consider composite effect measures, including the all-cause hazard ratio, that summarize patients’ outcomes as measured by several endpoints. [37] Future research could extend our research to composite effect measures.
	\item Even once an effect measure-outcome combination is chosen, different techniques to estimate the effect measure (e.g., maximum likelihood and Mantel-Haenszel) and different overarching approaches (e.g., stratification and product terms) may lead to opposite conclusions about the presence and direction of effect-measure modification. [38, 39] Existing literature demonstrates this for RR; future research could investigate this for other effect measure-outcome combinations.
	\item Warn et al. give Bayesian techniques for finding the posterior distributions of RR, OR, and RD. Future research could extend these techniques to RR*, HR, and HR*. From there, it could find the joint posterior distribution of the relative risk ratios $\frac{p2p3}{p1p4} \text{ and } \frac{(1 - p1)(1 - p4)}{(1 - p2)(1 - p3)}$. If the confidence region of this distribution lives entirely in the (<1, <1) or (>1, >1) quadrant, the prior distribution and new evidence together inform the conclusion of effect-measure modification on the \{RR, RR*, HR, HR*, RD, OR\} scales in the same direction.
	\item Since the two relative risks are concordant with Cheng’s preventative and generative causal powers, existing research [40] relating the causal powers to Bayesian networks could be readily extended to the two relative risks.
	\item Huitfeldt et al. show confounding and monotonicity assumptions for reaching counterfactual interpretations of the two relative risks and their reciprocals. Further research could, given those assumptions, assess the possibility and frequency of disagreement between these counterfactual outcome state transition (COST) parameters and non-COST effect measures (HR, HR*, RD, OR). [43]
\end{itemize}

\bibliographystyle{unsrt}
%\bibliography{EMMreferences5jan2020.bib}

\section*{Appendix A}
\label{AppendixA}
From Figure \ref{CommonVenn}, we conjectured that the probability that the two relative risks agree — or equivalently by our theorem in Section \ref{Theorem}, that all our effect measures agree — is $\frac{5}{6}$ when risks are randomly sampled from the (0,1) uniform distribution. In this appendix, we prove that conjecture.

\subsection*{Analytic proof that when RR and RR* agree, RD agrees with them}

In Section \ref{Theorem}, we proved algebraically that if the two relative risks agree, then the entire set of our effect measures \{RR, RR*, HR, HR*, RD, OR\} agrees. In this section, we provide an analytic proof that when the two relative risks agree, the risk difference agrees with them. We will use the framework of this proof in our proof that the probability of such agreement is $\frac{5}{6}$ when risks are randomly sampled from the uniform $(0,1)$ distribution.

As in Section \ref{Theorem}, let population $P$ describe the $(p1, p2)$ stratum, and let population $Q$ describe the $(p3, p4)$ stratum. For each effect measure EM, given risks $p_1, p_2, p_3 \in (0,1)$, we define $p_4^{*\text{EM}}$ to be the \textit{critical} value of $p_4$ at which the EM effect measure would indicate that the treatment or exposure affects populations $P$ and $Q$ equally. Notably, $p_4^{*\text{RR}} = \frac{p2p3}{p1}$, $p_4^{*\text{RD}} = p2 + p3 - p1$, and $p_4^{*\text{RR*}} = 1 - \frac{(1 - p2)(1 - p3)}{1 - p1}$.

The importance of $p4$* is that two effect measures disagree if (and only if) the true value of $p4$ falls between $p4$* for each effect measure. We will show this for $p2 > p1$ and $p4 > p3$. Let EM and FM be any two effect measures. Suppose that $p_4^{*\text{EM}} < p4 < p_4^{*\text{FM}}$. The left inequality gives that the EM effect measure considers population $Q$ to respond to treatment or exposure more strongly than population $P$, since $p4$ is more than what it would be were there no EM modification. Similarly, the right inequality gives that the FM effect measure considers population $P$ to respond to treatment or exposure more strongly than population $Q$. Hence EM and FM disagree as to the direction of effect-measure modification.

\begin{table}[ht]
    \centering
    \begin{tabular}{lllllll}
    Example & $p_1$ & $p_2$ & $p_3$ & $p_4^{*RR}$ & $p_4^{*RD}$ & $p_4^{*RR^*}$ \\
    A & 0.1 & 0.2 & 0.3 & 0.6 & 0.4 & $0.38$ \\
    B & 0.2 & 0.1 & 0.3 & 0.15 & 0.2 & 0.21 \\
    C & 0.2 & 0.3 & 0.1 & 0.15 & 0.2 & 0.21 \\
    D & 0.3 & 0.1 & 0.2 & $0.67$ & 0 & $-0.03$ \\
    \end{tabular}
    \caption{We consider a representative example for each of the four regions of $(0,1)^3$ given by the $p1 = p2$ and $p1 = p3$ planes.}
    \label{tab:Regions}
\end{table}

We start by looking at the four examples in Table \ref{tab:Regions}.
For each example, the theorem holds for all $p_4 \in (0,1)$, because $p_4^{*RD}$ is between $p_4^{*RR}$ and $p_4^{*RR^*}$. The planes $p_1 = p_2$ and $p_1 = p_3$ divide $(0,1)^3$, the space of $(p_1, p_2, p_3)$, into four open regions. Let us call each region by the boldface of the example it contains. For example, $(p_1 = 0.7, p_2 = 0.5, p_3 = 0.8)$ is in region \textbf{B} as $p_1 > p_2$ and $p_1 < p_3$. Consider any $B_1 = (p_1, p_2, p_3)$ in region \textbf{B}. By the intermediate value theorem, if in $B_1$, unlike B, $p_4^{*RD} > p_4^{*RR^*}$ or $p_4^{*RD} < p_4^{*RR}$, then for all continuous paths from B to $B_1$, even those living entirely inside region \textbf{B}, there must exist a point $(p_1, p_2 < p_1, p_3 > p_1)$ at which $p_4^{*RD} = p_4^{*RR^*}$ or $p_4^{*RD} = p_4^{*RR}$, respectively. But by Lemma, there are no points in region \textbf{B}, or \textbf{A} or \textbf{C} or \textbf{D}, for which $p_4^{*RD} = p_4^{*RR^*}$ or $p_4^{*RD} = p_4^{*RR}$. Thus no region has any point $(p_1, p_2, p_3)$ contradicting the $p_4^{*RR} < p_4^{*RD} < p_4^{*RR^*}$ or $p_4^{*RR} > p_4^{*RD} > p_4^{*RR^*}$ character shown in that region's example, so the theorem holds for all points in each region. In the remaining cases, $p_1 = p_2$ or $p_1 = p_3$, so all effect measures agree intuitively from the relationship between $p_3$ and $p_4$, or $p_2$ and $p_4$, respectively. All $(p_1, p_2, p_3) \in (0,1)^3$ lie in one of the four regions or the $p_1 = p_2$ or $p_1 = p_3$ plane, so if $p_4 \in (0,1)$ enables agreement between RR and RR*, then RD will also agree as desired.

\subsubsection*{Lemma: (\texorpdfstring{$p_4^{*RD} = p_4^{*RR^*}$}{p4*RD = p4*RR*} or \texorpdfstring{$p_4^{*RD} = p_4^{*RR}$}{p4*RD = p4*RR}) implies (\texorpdfstring{$p_1 = p_2$}{p1 = p2} or \texorpdfstring{$p_1 = p_3$)}{p1 = p3}}

Proof: Algebra confirms this Lemma for $p_4^{*RD} = p_4^{*RR}$:
\[ p_2 + p_3 - p_1 = \frac{p_2p_3}{p_1} \]
\[ p_1p_2 + p_1p_3 - p_1^2 - p_2p_3 = 0 \]
\[ (p_1 - p_2)(p_1 - p_3) = 0 \]
\[ p_1 = p_2 \text{ or } p_1 = p_3 \]

And similarly for $p_4^{*RD} = p_4^{*RR^*}$:
\[ p_2 + p_3 - p_1 = 1 - \frac{(1 - p_2)(1 - p_3)}{1 - p_1} \]
\[ (1 - p_1)(p_2 + p_3 - p_1) = 1 - p_1 - (1 - p_2)(1 - p_3) \]
\[ p_1^2 - (p_2 + p_3 + 1)p_1 + p_2 + p_3 = -p_1 + p_2 + p_3 - p_2p_3 \]
\[ p_1^2 - (p_2 + p_3)p_1 + p_2p_3 = 0 \]
\[ (p_1 - p_2)(p_1 - p_3) = 0 \]
\[ p_1 = p_2 \text{ or } p_1 = p_3 \]

Therefore, if $p_4^{*RD} = p_4^{*RR^*}$ or $p_4^{*RD} = p_4^{*RR}$, then $ p_1 = p_2 \text{ or } p_1 = p_3 $ as desired.

\subsection*{Theorem: Let \texorpdfstring{$p1, p2, p3, p4$}{p1, p2, p3, p4} be independent random variables each following a uniform (0,1) distribution. The probability that the set of effect measures \{RR, RR*, HR, HR*, RD, OR\} agrees is 5/6.}

Proof: By our theorem in Section \ref{Theorem}, it suffices to show that the probability of RR and RR* disagreeing is $\frac{1}{6}$. For each point in $(0,1)^3$, the conditional probability that RR and RR* disagree is the probability that $p4$ falls between $p4^{*RR}$ and $p4^{*RR^*}$. Since $p4$ takes a $(0,1)$ uniform distribution, this probability is $\min \{1, \max \{p4^{*RR}, p4^{*RR^*}\}\} - \max \{0, \min \{p4^{*RR}, p4^{*RR^*}\}\}$. Hence the overall probability $P$ that RR and RR* disagree is $\iiint_{(0,1)^3} (\min \{1, \max \{p4^{*RR}, p4^{*RR^*}\}\} - \max \{0, \min \{p4^{*RR}, p4^{*RR^*}\}\}) dp_{1,2,3}$. We partition $(0,1)^3$ into regions \textbf{A}, \textbf{B}, \textbf{C}, and \textbf{D} as above. This gives $P = $
\begin{gather*}
\iiint_{\textbf{A}} (\min \{1, \max \{p4^{*RR}, p4^{*RR^*}\}\} - \max \{0, \min \{p4^{*RR}, p4^{*RR^*}\}\}) \,dp_{1,2,3}
\\
+ \iiint_{\textbf{B}} (\min \{1, \max \{p4^{*RR}, p4^{*RR^*}\}\} - \max \{0, \min \{p4^{*RR}, p4^{*RR^*}\}\}) dp_{1,2,3}
\\
+ \iiint_{\textbf{C}} (\min \{1, \max \{p4^{*RR}, p4^{*RR^*}\}\} - \max \{0, \min \{p4^{*RR}, p4^{*RR^*}\}\}) dp_{1,2,3}
\\
+ \iiint_{\textbf{D}} (\min \{1, \max \{p4^{*RR}, p4^{*RR^*}\}\} - \max \{0, \min \{p4^{*RR}, p4^{*RR^*}\}\}) dp_{1,2,3}
\end{gather*}
Each of these integrals evaluates to $\frac{1}{24}$. We will compute the first integral; we leave the rest to the reader.

We know that the following are always true in region \textbf{A}:
\begin{itemize}
    \item $p1 < p2$ (by definition of region \textbf{A})
    \item $p1 < p3$ (by definition of region \textbf{A})
    \item $p4^{*RR} > p4^{*RR^*} \geq 0$.
        Earlier, we used the intermediate value theorem to show that the above imply $p4^{*RR} > p4^{*RD} > p4^{*RR^*}$. Furthermore, $p4^{*RR^*} < 0$ would imply $p3 < 1 - \frac{1 - p1}{1 - p2}$, an impossibility in region \textbf{A} since $p1 < p2$.
\end{itemize}
Hence we can resolve the extrema in our integral:
\begin{gather*}
    \iiint_{\textbf{A}} (\min \{1, \max \{p4^{*RR}, p4^{*RR^*}\}\} - \max \{0, \min \{p4^{*RR}, p4^{*RR^*}\}\}) \,dp_{1,2,3}
    \\
    = \iiint_{\textbf{A}} (\min \{1, p4^{*RR}\} - p4^{*RR^*}) \,dp_{1,2,3}
\end{gather*}
To resolve the remaining minimum, we will separately integrate the subregions in which each  candidate is the minimum:
\begin{gather*}
    \int_0^1 \int_{p1}^1 \int_{p1}^1 (\min \{1, \frac{p2 p3}{p1}\} - (1 - \frac{(1 - p2)(1 - p3)}{1 - p1})) \,dp3 \,dp2 \,dp1
    \\
    = \int_0^1 \int_{p1}^1 \int_{p1}^{\frac{p1}{p2}} \frac{p2 p3}{p1} \,dp3 \,dp2 \,dp1
    \\
    + \int_0^1 \int_{p1}^1 \int_{\frac{p1}{p2}}^1 1 \,dp3 \,dp2 \,dp1
    \\
    - \int_0^1 \int_{p1}^1 \int_{p1}^1 (1 - \frac{(1 - p2)(1 - p3)}{1 - p1}) \,dp3 \,dp2 \,dp1
    \\
    = \frac{1}{16} + \frac{1}{4} - \frac{13}{48} = \frac{1}{24}
\end{gather*}

Our integration over region \textbf{A} shows that the probability that $p1 < p2$, $p1 < p3$, and the two relative risks disagree is $\frac{1}{24}$. Similar integration over regions \textbf{B}, \textbf{C}, and \textbf{D} shows that probability to be $\frac{1}{24}$ for each of the other three inequality cases. Hence the overall probability that the two relative risks disagree is $\frac{1}{6}$. Applying our Theorem from Section \ref{Theorem}, we see that all our effect measures \{RR, RR*, HR, HR*, RD, OR\} agree with probability $\frac{5}{6}$ as desired.

\section*{Appendix B}
\label{AppendixB}
In this appendix, we show that if $0 < p4 < p2 < 1 < \frac{1 - p1}{1 - p2} < \frac{1 - p3}{1 - p4}$, then $\frac{\log(1 - \tilde{p_2}\text{RR*}_P)}{\log(1 - \tilde{p_2})} < \frac{\log(1 - \tilde{p}_4\text{RR*}_P)}{\log(1 - \tilde{p}_4)} < \frac{\log(1 - \tilde{p_4}\text{RR*}_Q)}{\log(1 - \tilde{p_4})}$.

\subsection*{Proof of \texorpdfstring{$\frac{\log(1 - \tilde{p_4}\text{RR*}_P)}{\log(1 - \tilde{p_4})} < \frac{\log(1 - \tilde{p_4}\text{RR*}_Q)}{\log(1 - \tilde{p_4})}$}{The Right Inequality}}
By assumption, RR*\textsubscript{P} $<$ RR*\textsubscript{Q}. Equivalently, $\log(1 - \tilde{p_4}\text{RR*}_P) > \log(1 - \tilde{p_4}\text{RR*}_Q)$. We now divide both sides by the negative value $\log p4 = \log(1 - \tilde{p_4})$, giving $\frac{\log(1 - \tilde{p_4}\text{RR*}_P)}{\log(1 - \tilde{p_4})} < \frac{\log(1 - \tilde{p_4}\text{RR*}_Q)}{\log(1 - \tilde{p_4})}$ as desired.

\subsection*{Proof of \texorpdfstring{$\frac{\log(1 - \tilde{p_2}\text{RR*}_P)}{\log(1 - \tilde{p_2})} < \frac{\log(1 - \tilde{p_4}\text{RR*}_P)}{\log(1 - \tilde{p_4})}$}{The Left Inequality}}

Let $x$ be such that $\tilde{p_2} \leq x \leq \tilde{p_4}$. Since $0 < x < 1$, we have $-\frac{x^2}{2} - \frac{x^3}{3} - \ldots < 0$. Equivalently, $-x - \frac{x^2}{2} - \frac{x^3}{3} - \ldots < -x = \frac{x^2 - x}{1 - x}$. The left side is the power series for $\log(1-x)$, so $\log(1 - x) < \frac{x^2 - x}{1 - x}$ and $x + (1 - x)\log(1 - x) < x^2$. Rearranging, $-\frac{x}{1 - x} - \log(1 - x) > - \frac{x^2}{1 - x}$ and $\frac{\frac{-x}{1 - x} - \log(1 - x)}{x^2} + \frac{1}{1 - x} > 0$. But the left side is the derivative of $\frac{\log(1 - x)}{x} - \log(1 - x)$, so that expression must be strictly increasing over the open unit interval (as $0 < x < 1$ was the only property of $x$ we used so far). In particular, $x < x\text{RR*}_P = \frac{x\tilde{p_1}}{\tilde{p_2}} \leq \frac{\tilde{p_4}\tilde{p_1}}{\tilde{p_2}} < \tilde{p_3} < 1$, so $\frac{1 - x}{x}\log(1 - x) < \frac{1 - x\text{RR*}_P}{x\text{RR*}_P}\log(1 - x\text{RR*}_P)$. Rearranging, $\frac{(1 - x\text{RR*}_P)\log(1 - x\text{RR*}_P)}{(1 - x)\log(1 - x)} < \text{RR*}_P$ and $\frac{\log(1 - x\text{RR*}_P)}{1 - x} - \frac{\text{RR*}_P\log(1 - x)}{1 - x\text{RR*}_P} > 0$. Dividing both sides by the positive $(\log(1 - x))^2$, the left side becomes the derivative of $\frac{\log(1 - x\text{RR*}_P)}{\log(1 - x)}$. Hence that expression is strictly increasing over $\tilde{p_2} \leq x \leq \tilde{p_4}$. As desired, $\frac{\log(1 - \tilde{p_2}\text{RR*}_P)}{\log(1 - \tilde{p_2})} < \frac{\log(1 - \tilde{p_4}\text{RR*}_P)}{\log(1 - \tilde{p_4})}$.

\section*{Appendix C}
\label{Appendix C}
In this appendix, we provide a remedial measure for defining the cumulative hazard ratios when the equal follow-up periods assumption fails.

If the assumption of equal follow-up periods does not hold, it may be possible to define the cumulative hazard ratios by letting $t_f$ represent the lesser of the durations of the follow-up periods. If the proportional hazards (recovery rates) assumption holds, then the interpretation of HR (HR*) seen in our case studies (Sections \ref{HCV}, \ref{Melanoma}, and \ref{COVID}) holds during the shorter follow-up period. However, HR (HR*) would no longer consider all participants reaching (recovering from) the outcome, so we would no longer have HR $= \frac{\log(1 - p2)}{\log(1 - p1)}$ (HR* = $\frac{\log p1}{\log p2}$) in general. It would then be possible for HR (HR*) to disagree with other effect measures even when RR and RR* agree.

\section*{Appendix D}
\label{AppendixD}
\begin{figure}[htbp]
\centering
\includegraphics[scale = 0.425]{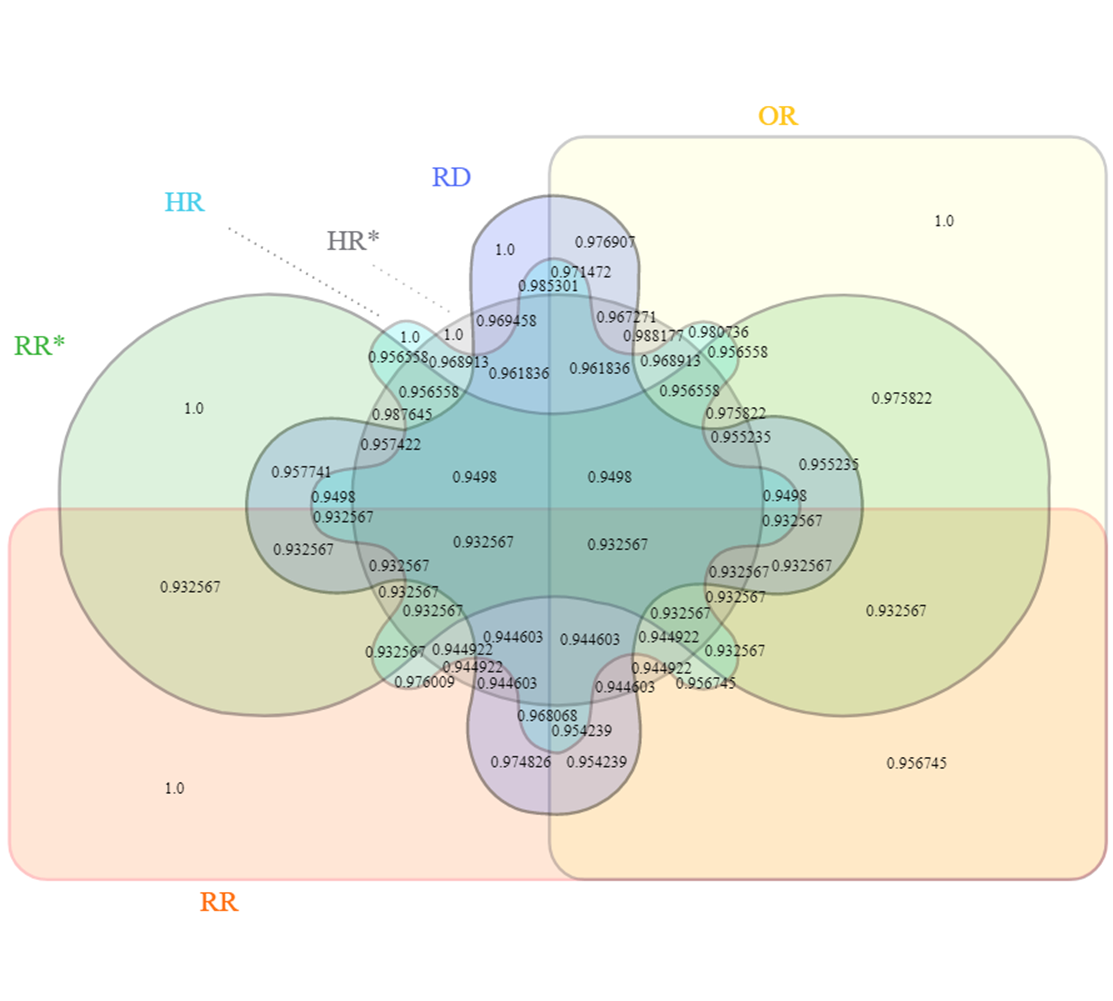}
\caption{We present the approximate probability of agreement for various sets of effect measures when control risks $p1$ and $p3$ are randomly sampled from the (0,1) distribution, while treatment/exposure risks $p2$ and $p4$ are sampled from the asymmetric tent distributions achieving respective maxima at $p1$ and $p3$. In this way, $p2$ and $p4$ positively correlated with $p1$ and $p3$.}
\label{TentVenn}
\end{figure}

Our Monte Carlo simulation in Section \ref{MonteCarlo} gave equal consideration to all possible values $(p1, p2, p3, p4)$ by independently sampling each risk from the uniform $(0,1)$ distribution. In practice, there is generally a correlation between patients' potential outcomes, leading $p2$ to depend on $p1$ and $p4$ to depend on $p3$. [45] While $(p1 = 0.1, p2 = 0.9, p3 = 0.8, p4 = 0.3)$ could be the outcome of a two-strata binary outcome randomized control trial, $(p1 = 0.56, p2 = 0.53, p3 = 0.78, p4 = 0.74)$ would be a much more likely outcome given no additional context. Unfortunately, there is no standard for modeling dependence: While we intuitively believe $(0.56, 0.53, 0.78, 0.74)$ to be more likely than $(0.1, 0.9, 0.8, 0.3)$, we have no basis for specifying how much more likely for simulation purposes. Therefore, Section \ref{MonteCarlo} and Appendix A proceed with independent samples to show the theoretical abundance of quadruples $(p1, p2, p3, p4)$ for which RR and RR* disagree, while our case studies (Sections \ref{HCV}, \ref{Melanoma}, \ref{COVID}) show real-world examples living in this theoretical abundance.

In this appendix, we reperform our Monte Carlo simulation by sampling $p2$ and $p4$ from tent distributions with respective maxima at $p1$ and $p3$. For example, the quadruple $(0.56, 0.53, 0.78, 0.74)$ is roughly 22 times as likely to be simulated as the quadruple $(0.1, 0.9, 0.8, 0.3)$. More precisely, let $f_2,f_4 : [0,1] \rightarrow \mathbb{R}$ be the probability density functions governing $p2$ and $p4$. As in the original simulation, we assign $f_2(0) = f_4(0) = f_2(1) = f_4(1) = 0$. To encode the dependence of $p2$ on $p1$ and $p4$ on $p3$, we want $f_2$ to achieve a maximum at $p1$ and $f_4$ at $p3$. Of the infinitely many probability density functions with these properties, we choose the asymmetric tent map with $\mu = 2$ to linearly ``connect the dots." We get the Venn Diagram presented in Figure \ref{TentVenn}. We see that the estimated probability of disagreement under this model of dependence is only 0.067433. Disagreement is still probable enough to motivate the reporting of underlying risks or multiple effect measures.

In the accompanying code, we additionally allow the user to specify lower and upper bounds for the risks. For a lower bound $L$ and an upper bound $U$, this gives the cumulative distribution function
\begin{equation*}
    F(p2) = \begin{cases}
        \frac{p2^2 - 2Lp2 + L^2}{(p1 - L)(U - L)} & p2 < p1 \\
        \frac{p2^2 + p1(L - 2p1) - U(2p2 + L - 3p1)}{(U - L)(U - p1)} & p2 \geq p1
    \end{cases}.
\end{equation*}

%\bibliographystyle{unsrtnat}
%\bibliography{references}  %%% Uncomment this line and comment out the ``thebibliography'' section below to use the external .bib file (using bibtex) .


\begin{thebibliography}{99}
\bibitem{journal-1} Brumback B., Berg A., On effect-measure modification: Relationships among changes in the relative risk, odds ratio, and risk difference, Stat. Med., 2008, 27(18), 3453-3465.
\bibitem{journal-2} Chu H., Nie L., Chen Y., Huang Y., Sun W., Bivariate random effects models for meta-analysis of comparative studies with binary outcomes: Methods for the absolute risk difference and relative risk, Stat. Methods Med. Res., 2012, 21(6), 621-633.
\bibitem{journal-3} Flandre P., Choice of effect measure in HIV randomized trials, AIDS, 2015, 29(15), 2057-2060.
\bibitem{journal-4} VanderWeele T.J., Confounding and Effect Modification: Distribution and Measure, Epidemiologic Methods, 2015, 1(1), 55-82.
%Technical report
\bibitem{journal-5} Wodtke G., Almirall D., Estimating Heterogeneous Causal Effects with Time-Varying Treatments and Time-Varying Effect Moderators, Population Studies Center, 2015, Report 15-839.
\bibitem{journal-6} Wodtke G.T., Almirall D., Estimating Heterogeneous Causal Effects with Time-Varying Treatments and Time-Varying Effect Moderators: Structural Nested Mean Models and Regression with Residuals. Sociological Methodology, 2017, 47(1), 212-245.
\bibitem{journal-7} Hossin M.Z., {\"O}stergren O., Fors S., Is the Association Between Late Life Morbidity and Disability Attenuated Over Time? Exploring the Dynamic Equilibrium of Morbidity Hypothesis, The Journals of Gerontology: Series B, 2019, 74(8), e97-e106.
\bibitem{journal-8} Greenland S., Multiple comparisons and association selection in general epidemiology, International Journal of Epidemiology, 2008, 37(3), 430-434.
%Technical report
\bibitem{journal-9} Phillippo D., Ades T., Dias S., Palmer S., Abrams K.R., Welton N., NICE DSU Technical Support Document 18: Methods for population-adjusted indirect comparisons in submissions to NICE, NICE Decision Support Unit, 2016, Report 18.
\bibitem{journal-10} Dahabreh I.J., Hayward R., Kent D.M., Using group data to treat individuals: understanding heterogeneous treatment effects in the age of precision medicine and patient-centred evidence, International Journal of Epidemiology, 2016, 45(6), 2184-2193.
\bibitem{journal-11} Almirall D., Griffin B.A., McCaffrey D.F., Ramchand R., Yuen R.A., Murphy S.A., Time-varying effect moderation using the structural nested mean model: estimation using inverse-weighted regression with residuals, Stat. Med., 2014, 33(20), 3466-3487.
%5 pages long, but journal site says only page 1248.
\bibitem{journal-12} Schwartz L.M., Woloshin S., Dvorin E.L., Welch H.G., Ratio measures in leading medical journals: structured review of accessibility of underlying absolute risks, BMJ, 2006, 333(7581), 1248.
\bibitem{journal-13} Scanlan J., The Mismeasure of Health Disparities, Journal of Public Health Management and Practice, 2016, 22(4), 415-419.
\bibitem{journal-14} Furuya-Kanamori L., Doi S.A.R., The outcome with higher baseline risk should be selected for relative risk in clinical studies: a proposal for change to practice, Journal of Clinical Epidemiology, 2014, 67(4), 364-367.
\bibitem{proceedings} Scanlan J., Measurement Problems in the National Healthcare Disparities Report, Paper presented at American Public Health Association 135th Annual Meeting \& Exposition, (Washington D.C., United States), 2007, November 3-7
\bibitem{journal-15} Orenstein W.A., Bernier R.H., Dondero T.J., Hinman A.R., Marks J.S., Bart K.J., Sirotkin B., Field evaluation of vaccine efficacy, Bulletin of the World Health Organization, 1985, 63(6), 1055-1068.
\bibitem{journal-16} Cheng P.W., From covariation to causation: A causal power theory, Psychological Review, 1997, 104(2), 367-405.
\bibitem{journal-17} Hiddleston E., Causal Powers, British J. Philos. Sci., 2005, 56(1), 27-59.
\bibitem{journal-18} Pearl J., Probabilities of Causation: Three Counterfactual Interpretations and their Identification, Synthese, 1999, 121(1-2), 93-149.
\bibitem{journal-19} Edwards A.W.F., The Measure of Association in a 2 $\times$ 2 Table. J. Roy. Statist. Soc. Ser. A, 1963, 126(1), 109-114.
\bibitem{journal-20} Hern{\'a}n, M.A., The Hazards of Hazard Ratios, Epidemiology, 2010, 21(1), 13-15.
\bibitem{journal-21} Dahari H., Canini L., Graw F., Uprichard S.L., Araujo E.S.A., Penaranda G., Coquet E., Chiche L., Riso A., Renou C., Bourliere M., Cotler S.J., Halfon P., HCV kinetic and modeling analyses indicate similar time to cure among sofosbuvir combination regimes with daclatasvir, simeprevir or ledipasvir, Journal of Hepatology, 2016, 64(6), 1232-1239.
\bibitem{journal-22} GBD 2015 Disease and injury Incidence and Prevalence Collaborators, Global regional, and national incidence, prevalence, and years lived with disability for 310 diseases and injuries, 1990-2015: a systematic analysis for the Global Burden of Disease Study 2015, Lancet, 2016, 388(10053), 1545-1602.
\bibitem{journal-23} Ramsey S., Blough D., Kirchhoff A., Kreizenbeck K., Fedorenko C., Snell K., Newcomb P., Hollingworth W., Overstreet K., Washington State Cancer Patients Found To Be At Greater Risk For Bankruptcy Than People Without A Cancer Diagnosis, Health Affairs, 2013, 32(6), 1143-1152.
\bibitem{journal-24} Mahase E., Covid-19: death rate is 0.66\% and increases with age, study estimates, BMJ, 2020, 369.
\bibitem{journal-25} Winnie S., Kaplan R., Why Do Countries' COVID-19 Death Rates Vary So Much?, MedPage Today, 2020.
\bibitem{journal-26} Kenyon C., Flattening-the-curve associated with reduced COVID-19 case fatality rates- an ecological analysis of 65 countries, Journal of Infection, 2020, 81(1), e98-e99.
%technical report
\bibitem{journal-27} The COVID-19 Task force of the Department of Infectious Diseases and the IT Service, Integrated surveillance of COVID-19 in Italy, Istituto Superiore di Sanit{\'a}, 2020, Ordinanza 640. \url{https://www.epicentro.iss.it/en/coronavirus/bollettino/Infografica_3giugno\%20ENG.pdf}.
%technical report
\bibitem{journal-28} CONACYT, COVID-19 Tablero M{\'e}xico, 2020. \url{http://datos.covid-19.conacyt.mx/index.php}.
\bibitem{newspaper-1} Luhnow D., de C{\'o}rdoba J., As Covid-19 Hits Developing Countries, Its Victims Are Younger, 2020, June 19, WSJ.
\bibitem{journal-29} Mackenbach, J.P., Should we aim to reduce relative or absolute inequalities in mortality?, European Journal of Public Health, 2015, 25(2), 185.
\bibitem{journal-30} Bauld L., Day P., Judge K., Off Target: A Critical Review of Setting Goals for Reducing Health Inequalities in the United Kingdom, International Journal of Health Services, 2008, 38(3), 439-454.
\bibitem{journal-31} VanderWeele T.J., Knol M.J., A Tutorial on Interaction, Epidemiologic Methods, 2014, 3(1), 33-72.
\bibitem{journal-32} Heberle H., Meirelles G.V., da Silva F.R., Telles G.P., Minghim R., InteractiVenn: a web-based tool for the analysis of sets through Venn diagrams, BMC Bioinformatics, 2015, 16(1), 169.
\bibitem{journal-33} Greenland S., Model-based Estimation of Relative Risks and Other Epidemiologic Measures in Studies on Common Outcomes and in Case-Control Studies, American Journal of Epidemiology, 2004, 160(4), 301-305.
\bibitem{journal-34} Gilbert P.B., Blette B.S., Shepherd B.E., Hudgens M.G., Post-randomization Biomarker Effect Modification Analysis in an HIV Vaccine Clinical Trial, Journal of Causal Inference, 2020, 8(1), 54-69.
\bibitem{journal-35} Rauch G., Jahn-Eimermacher A., Brannath W., Kieser M., Opportunities and challenges of combined effect measures based on prioritized outcomes, Stat. Med., 2014, 33(7), 1104-1120.
\bibitem{journal-36} Sturmer T., Rothman K.J., Glynn R.J., Insights into different results from different causal contrasts in the presence of effect-measure modification, Pharmacoepidemiology and Drug Safety, 2006, 15(10), 698-709.
\bibitem{journal-37} Buckley J.P., Doherty B.T., Keil A.P., Engel S.M., Statistical Approaches for Estimating Sex-Specific Effects in Endocrine Disruptors Research, Environmental Health Perspectives, 2017, 125(6), 067013.
\bibitem{journal-38} Hongjing L., Yuille A.L., Lijeholm M., Cheng P.W., Holyoak K.J., Bayesian Generic Priors for Causal Learning, Psychological Review, 2008, 115(4), 955-984.
\bibitem{journal-39} Sheps M., Shall We Count the Living or the Dead? New England Journal of Medicine, 1958, 259(25), 1210-1214.
\bibitem{journal-40} Deeks J., Issues in the selection of a summary statistic for meta-analysis of clinical trials with binary outcomes, Statistics in Medicine, 2002, 21(11), 1575-1600.
\bibitem{journal-41} Huitfeldt A., Goldstein A., Swanson S.A., The choice of effect measure for binary outcomes: Introducing counterfactual outcome state transition parameters, Epidemiologic methods, 2018, 7(1).
\bibitem{arxiv-1} Baker R., Jackson D., A new measure of treatment effect for random-effects meta-analysis of comparative binary outcome data, arXiv:1806.03471 [stat]
\bibitem{journal-42} Huitfeldt A., Odds ratios are not conditional risk ratios, Journal of Clinical Epidemiology, 2017, 84, 191.
\end{thebibliography}




% \end{thebibliography}

\end{document}